\documentclass[10pt,twocolumn,oneside,letterpaper,journal]{IEEEtran}
\ifCLASSINFOpdf
  \usepackage[pdftex]{graphicx}
  \graphicspath{{../}}
  \DeclareGraphicsExtensions{.pdf,.jpg,.png}
\else
\fi

\usepackage{cite}
\usepackage{xcolor}
\usepackage{pgfplots}
\usepackage{psfrag}
\usepackage{color}

\usepackage[cmex10]{amsmath}
\usepackage{array}
\usepackage{mdwmath}
\usepackage{mdwtab}
\usepackage{eqparbox}
\usepackage{stfloats}
\usepackage[caption=false,font=footnotesize]{subfig}

\usepackage[colorlinks=true,bookmarks=true,linkcolor=black,anchorcolor=green,citecolor=black]{hyperref}

\usepackage{booktabs}
\usepackage{textcomp}
\usepackage{colortbl}

\begin{document}
%
% paper title
% can use linebreaks \\ within to get better formatting as desired
\title{In-vivo Network of Sensors and Actuators}
%
%
% author names and IEEE memberships
% note positions of commas and nonbreaking spaces ( ~ ) LaTeX will not break
% a structure at a ~ so this keeps an author's name from being broken across
% two lines.
% use \thanks{} to gain access to the first footnote area
% a separate \thanks must be used for each paragraph as LaTeX2e's \thanks
% was not built to handle multiple paragraphs
%
\bstctlcite{IEEEexample:BSTcontrol}

%===========================================================================================================
%                                                  Authors
%===========================================================================================================
\author{Mo~Zhao
        and~Robert~H.~Blick% <-this % stops a space
%\author{Mo~Zhao,~\IEEEmembership{Student Member,~IEEE,}
%        and~Robert~H.~Blick,~\IEEEmembership{Member,~IEEE}% <-this % stops a space
% \thanks{Manuscript received July 30, 2010; revised October 11, 2010.}% <-this % stops a space
\thanks{The authors are with the Department of Electrical and Computer Engineering, University of Wisconsin-Madison, Madison,
WI 53706, USA, and Institute for Applied Physics, University of Hamburg, Jungiusstrasse 11c, 20355 Hamburg, Germany. Corresponding e-mail: mzhao8@wisc.edu.}
}

\maketitle

%===========================================================================================================
%                                                  Abstract
%===========================================================================================================
\begin{abstract}
%\boldmath
An advanced system of sensors/actuators should allow the direct feedback of a sensed signal into an actuation, e.g., an action potential propagation through an axon or a special cell activity might be sensed and suppressed by an actuator through voltage stimulation or chemical delivery. Such a complex procedure of sensing and stimulation calls for direct communication among these sensors and actuators. In addition, minimizing the sensor/actuator to the size of a biological cell can enable the cell-level automatic therapy. For this objective, we propose such an approach to form a peer-to-peer network of \emph{in vivo} sensors/actuators (S/As) that can be deployed with or even inside biological cells. The S/As can communicate with each other via electromagnetic waves of optical frequencies. In comparison with the comparable techniques including the radio-frequency identification (RFID) and the wireless sensor network (WSN), this technique is well adapted for the cell-level sensing-actuating tasks considering the requirements on size, actuation speed, signal-collision avoidance, etc.

\end{abstract}
% IEEEtran.cls defaults to using nonbold math in the Abstract.
% This preserves the distinction between vectors and scalars. However,
% if the journal you are submitting to favors bold math in the abstract,
% then you can use LaTeX's standard command \boldmath at the very start
% of the abstract to achieve this. Many IEEE journals frown on math
% in the abstract anyway.

%===========================================================================================================
%                                                  Key Words
%===========================================================================================================
% Note that keywords are not normally used for peerreview papers.
\begin{IEEEkeywords}
biosensors, body area networks, cancer therapy, optical antenna arrays, communication protocols.
\end{IEEEkeywords}

% For peer review papers, you can put extra information on the cover
% page as needed:
% \ifCLASSOPTIONpeerreview
% \begin{center} \bfseries EDICS Category: 3-BBND \end{center}
% \fi

% For peerreview papers, this IEEEtran command inserts a page break and
% creates the second title. It will be ignored for other modes.
\IEEEpeerreviewmaketitle

%===========================================================================================================
%                                      The First Section (Introduction)
%===========================================================================================================
\section{Introduction}

% The very first letter is a 2 line initial drop letter followed
% by the rest of the first word in caps.
% form to use if the first word consists of a single letter:
% \IEEEPARstart{A}{demo} file is ....
%
% form to use if you need the single drop letter followed by
% normal text (unknown if ever used by IEEE):
% \IEEEPARstart{A}{}demo file is ....
%
% Some journals put the first two words in caps:
% \IEEEPARstart{T}{his demo} file is ....
%
% Here we have the typical use of a "T" for an initial drop letter
% and "HIS" in caps to complete the first word.
%
% You must have at least 2 lines in the paragraph with the drop letter
% (should never be an issue)

\IEEEPARstart{W}{ith} the rapid development of biological technology and nanotechnology, the current state-of-art sensors for \emph{in vivo} applications allow the extraction of information from biological cells, such as the action potential, the pH-value, the pharmacokinetic parameters, proteinase activity, and even the molecular events, by converting chemical information or conformational change into electrical or optical signals \cite{Burke2010,Plaxco2011,Eckert2013,Ferguson2013,Uusitalo2012}. In turn, \emph{in vivo} actuators can exert specific stimulations or adjust therapeutic dose on targeted cells. An advanced system of sensors/actuators should allow the direct feedback of a sensed signal into an actuation, e.g., an action potential propagation through an axon or a special cell activity might be sensed and suppressed by an actuator through voltage stimulation or chemical delivery. Such a complex procedure of sensing and stimulation calls for direct communication among these sensors and actuators. In addition, minimizing the sensor/actuator to the size of a biological cell can enable the cell-level automatic therapy, which is a dream of engineers and doctors \cite{Burke2010}.

In recent years, the wireless body area networks (WBANs), which interconnect tiny nodes with sensor or actuator capabilities in, on, or around a human body for short range communications, have been sustained attention \cite{Cao2009,Yuce2010,Seyedi2013,Chavez-Santiago2013}. To adapt the human body environments and reduce power consumption, in February 2012, IEEE published the first international standard for WBANs, IEEE Std 802.15.6 \cite{Seyedi2013,Chavez-Santiago2013,Zhang2014,IEEE15.6}. In a medical environment, WBANs may consist of wearable or implantable sensor nodes that sense biological information from the human body and transmit to a control device worn on the body or placed at an accessible location for ambulatory health/therapy status monitoring \cite{Cao2009,Yuce2010}. WBSNs' intra-body communication (IBC) is based mainly on capacitive coupling or galvanic coupling. Technical challenges were magnified for the IBC because of the lack of a corresponding mathematical model and the impossibility of conducting in-body measurements \cite{Seyedi2013,Chavez-Santiago2013,Zhang2014}. These hinder the deployment of WBANs to the applications of the cell-level automatic therapy.

In the past decade, there has been also a significant development in the cell-level cancer therapy. The localized surface plasmon resonance effects of gold nanoparticles, which strongly enhance the absorption from near-infrared (NIR) radiation, have been applied to photothermal cancer ablation, becoming a very perspective therapy path \cite{Maltzahn2009,Huang2006,Hirsch2003}. Similarly, single-walled carbon nanotubes (SWNTs), with their low cost, also show a strong optical absorbance in NIR spectral windows. This intrinsic optical properties, with SWNTs' functionalization and transporting capabilities, can be used for cancer cell position \cite{Iverson2013}, photothermal ablation \cite{Kam2005,Chakravarty2008,Zhou2009} and drug delivery \cite{Liu2008,Tripiscianoa2009}. In these novel therapy modes, the irradiation dose to destroy tumor cells relies on the absorption potency of nanoantennas. But the quantitative photothermal model for the in vivo absorption of plasmonic nanomaterials has widely remained absent from their testing \cite{Maltzahn2009}. This forms an obstacle to translation to effective clinical use.

In order to solve the problem of clinical use of these novel therapy modes, in this paper, we suggest a new localized body area network solution, which can achieve a precise drug or irradiation dose control through the close-loop feedback. This solution will be a power promotion to our idea about personalized medicine, with optimal doses for each patient to maximize efficacy and minimize side effects \cite{Plaxco2011,Ferguson2013}.

% An example of a floating figure using the graphicx package.
% Note that \label must occur AFTER (or within) \caption.
% For figures, \caption should occur after the \includegraphics.
% Note that IEEEtran v1.7 and later has special internal code that
% is designed to preserve the operation of \label within \caption
% even when the captionsoff option is in effect. However, because
% of issues like this, it may be the safest practice to put all your
% \label just after \caption rather than within \caption{}.
%
% Reminder: the "draftcls" or "draftclsnofoot", not "draft", class
% option should be used if it is desired that the figures are to be
% displayed while in draft mode.

\begin{figure*}[!t]
\centering
\includegraphics[width=14.0cm]{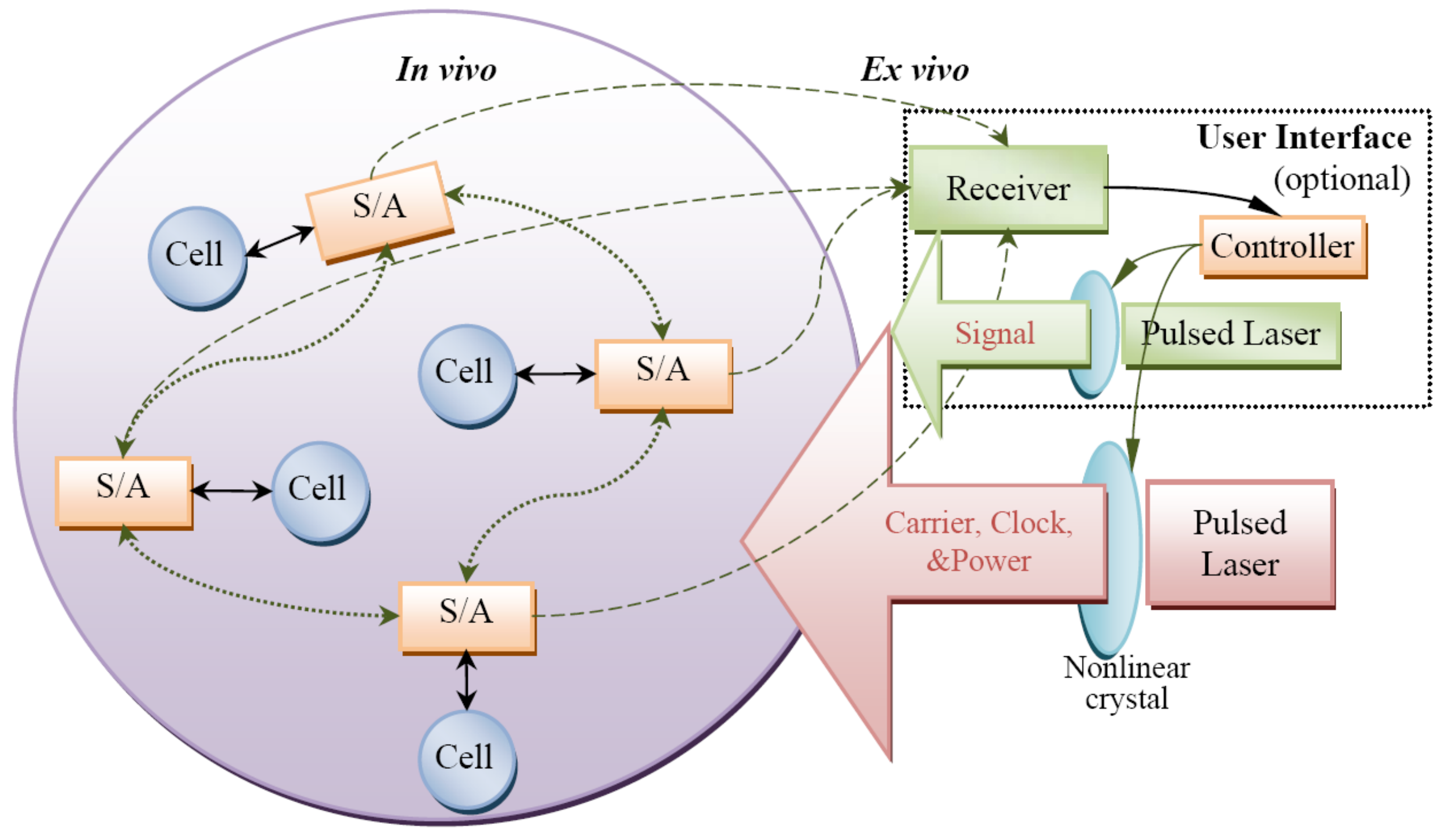}
% where an .eps filename suffix will be assumed under latex,
% and a .pdf suffix will be assumed for pdflatex; or what has been declared
% via \DeclareGraphicsExtensions.
\caption{A three-dimensional network of implanted sensors/actuators (S/As) which can intelligently interact with biological cells: A sensor extract the chemical information of a cell and converts it into electrical signal (solid line with arrows). Then it decides whether to excite the corresponding actuator by transmitting the commands (dotted line with arrows). In addition, a pulsed laser placed \emph{ex vivo} provides not only for the carrier of signals, but also for the power supply and the clock of all S/As. Thus, the peer-to-peer network of \emph{in vivo} S/As is formed. For flexibility of initialization, we can optionally add a `user interface', which merely communicates with S/As (dashed line with arrows) before the S/As work normally. The mode can be excited by using the controller of the user interface to block the clock signal for a long interval.}
\label{cell_fig1}
\end{figure*}

% Note that IEEE typically puts floats only at the top, even when this
% results in a large percentage of a column being occupied by floats.

% needed in second column of first page if using \IEEEpubid
%\IEEEpubidadjcol

We propose such an approach to form a layered peer-to-peer network of \emph{in vivo} sensors/actuators (S/As) that can be deployed with or even inside biological cells (see Fig. \ref{cell_fig1}). The S/As can communicate with each other via electromagnetic waves of optical frequencies. This carrier of signals is a sequence of ultrashort optical pulses provided by an \emph{ex vivo} pulsed laser. Meanwhile, the laser pulses also provide the power supply and the clock of all S/As. In order to better initialize the S/A network, we can add a `user interface' working only in the learning mode or for clock synchronization. It uses photodetectors \emph{ex vivo} to detect the optical signals transmitted from the S/As. The controller of the interface can block the laser beam for a long interval through the nonlinear optical crystal, acting as the signals for mode switching or synchronization. A second laser is operated at the frequency of S/A communication and the beam is coded with signals, but it is only optional for data input in the leaning mode. In comparison with the comparable techniques including the radio-frequency identification (RFID) \cite{EPC2008} the wireless sensor network (WSN) \cite{IEEE15.4}, and the wireless body area network (WBAN) \cite{IEEE15.6}, this technique is well adapted for the cell-level sensing-actuating tasks considering the requirements on size, communication range, actuation speed, signal-collision avoidance, synchronization mechanism, energy harvesting, etc. In the following, we discuss its novelties and how they contribute to this application.

%===========================================================================================================
\section{Device Minimization}

In order to implement cellular sensing/actuation, the total size of the S/As should be comparable to the size of biological cells, namely one to several microns. Smaller sizes could enable the implantation of S/As inside biological cells and hence would be even more attractive. This calls for structural minimization of not only the circuitry, but also the devices like the antenna and light source.

\textbf{\emph{Micron-Scale Optical Antenna:}}
How to minimize the antenna is an important issue for wireless devices, since the size of antennas has close relation with the available amount of power for utilization, even without consideration of the specific geometry. However, for the merely micron-scale area allowed by the applications of cellular sensing/actuation, the power of electromagnetic radiations can only provide a small measure of power. In this sense, we need to enhance the efficiency of power utilization. This calls for using the radiation with the wavelength comparable to the micron-scale antenna size and using the optical antenna which directly convert light to light. Here, the concept of optical antenna is rather a combination of the conventional optical antenna for light concentration and emission (which efficiently collects the energy of free-space radiation to a confined region of sub-wavelength size or vice versa). It first couples the optical far-field to the near-field so as to effectively interact with the nonlinear-optical metals, and it then transmits the near-field to the free space with intensity focused in some expected directions (i.e., radiation patterns) \cite{Taminiau2008,Zia2008}. Using appropriate structures, the extraordinarily enhanced emission and extremely small excitation volume would enable the study of single-biomolecule interactions \cite{Garcia-Parajo2008}. The frequency also changes, because the resonance of antenna filters out the redundant frequency components caused by the nonlinear effect. In our layout, the optical antenna should be composed of arrays of nanoantennas (e.g., holes in gold plane or gold particles), which can not only facilitate the interaction of the optical field and the plasmons, but also be controlled separately by applied voltages. Thus, the visible/near-infrared frequency assists to realize the minimization of antennas to the microscale.

\textbf{\emph{External Sources for Signals:}}
Another advantage of the optical antenna arrays is that they avoid the internal light source (e.g., LED or laser for common optoelectronic devices) which requires a high power supply and takes more space. Instead, we use a pulsed laser \emph{ex vivo} to produce the signal carrier (all `1' pulses), which is then coded and retransmitted (in another frequency channel and in the preferred direction) by the optical antenna arrays. Accordingly, the arrays simply convert the electrical signal (produced by the S/As) into light using the external light source. Nevertheless, the light signal can only travel for a short distance due to its low intensity, but it is appropriate for the peer-to-peer communication. In this sense, the ultrafast laser beam can enhance the sensitivity of photo-detection while keeping a low level of thermal effect.

\textbf{\emph{Synchronization by Laser Pulses:}}
The laser pulses serving as the signal carriers can also provide the clock for circuits, using a photodetector to convert them into electric pulses. First, this approach saves the resonator circuit and crystals to produce the clocks for S/As. Second, it synchronizes all the S/As, so that each can know the schedule of others and predict their working conditions using the same clock. Thus, we can organize their sequences using the same instruction cycle, like the internal parts of a computer. Third, it also saves the circuit for synchronization between the internal clock and the external clock when receiving or transmitting signals. Conventionally, the clock of a signal should be much slower than the internal clock of the receiver, in order to reduce the error of synchronization (namely the minimum cycle of the internal clock). However, in our case, the receiver and signal (from another S/A) are naturally synchronized and the conventional method to synchronize a fast clock and a slow clock becomes unnecessary.

\textbf{\emph{Simplification of Communication Protocols:}}
In order to reduce the scale of circuitry, we expect to minimize the complexity of the communication protocols. Unlike other comparable techniques including RFID \cite{EPC2008}, WSN \cite{IEEE15.4} and WBAN \cite{IEEE15.6}, which can employ more complex communication protocols and support longer communication distances, the \emph{in vivo} environment for cancer therapy has a completely different protocol requirements. On the one hand, the \emph{in vivo} environment's component capability is severely restricted by size, so that the complex communication protocols can not be supported. On the other hand, the \emph{in vivo} environment's networking requirement is relatively simple and do not needs to adopt a set of very complex communication protocols. Based these consideration, we have designed a new communication architecture suitable for the \emph{in vivo} application environment. In our application, only a few sorts of instructions/messages are transmitted, including the command from a sensor to an actuator, the acknowledge signal, the notifying/blocking signal, and extended messages such as for relays. Thus, in order to physically simplify the signals (to reduce the component's complexity) and also to shorten the transmission time (to reduce the component's energy consumptions and actuation delays), we perform a cross-layer design \cite{Lin2013,Fu2014}, which consider uniformly physical layer's implementation complexity and logic layers' functions needs. Hence, we optimize the signal-processing logic and the length of signal for transmission and reception, by uniformly setting the sequence of signals and the control bits. The concept of packet is no longer in mere logic, but defined by physical cycles of the laser pulses. In addition, the conventional CSMA/CA technique, widely used in the wireless communication \cite{IEEE15.4,IEEE11}, is no longer appropriate for our application because of its actuation delays, inefficiency, and increasing circuit complexity. Instead, we suggest a new approaches to deal with the signal collisions, Collision Detection when Muted and Exit when Colliding (CDWM/EWC). Further, we use the simplest way of signal coding: the pulse with power is denoted by `1', otherwise `no signal' is denoted by `0'. Compared to the communication protocols that usually adopt complex coding method (e.g., by varying the pulse duration and waveform), this coding is easy to realize in circuit and enables our new nonblocking collision management mechanism, CDWM/EWC.

%===========================================================================================================
\section{Communication Frequencies and Patterns}

\begin{figure*}[!t]
\centering
\includegraphics[width=14.0cm]{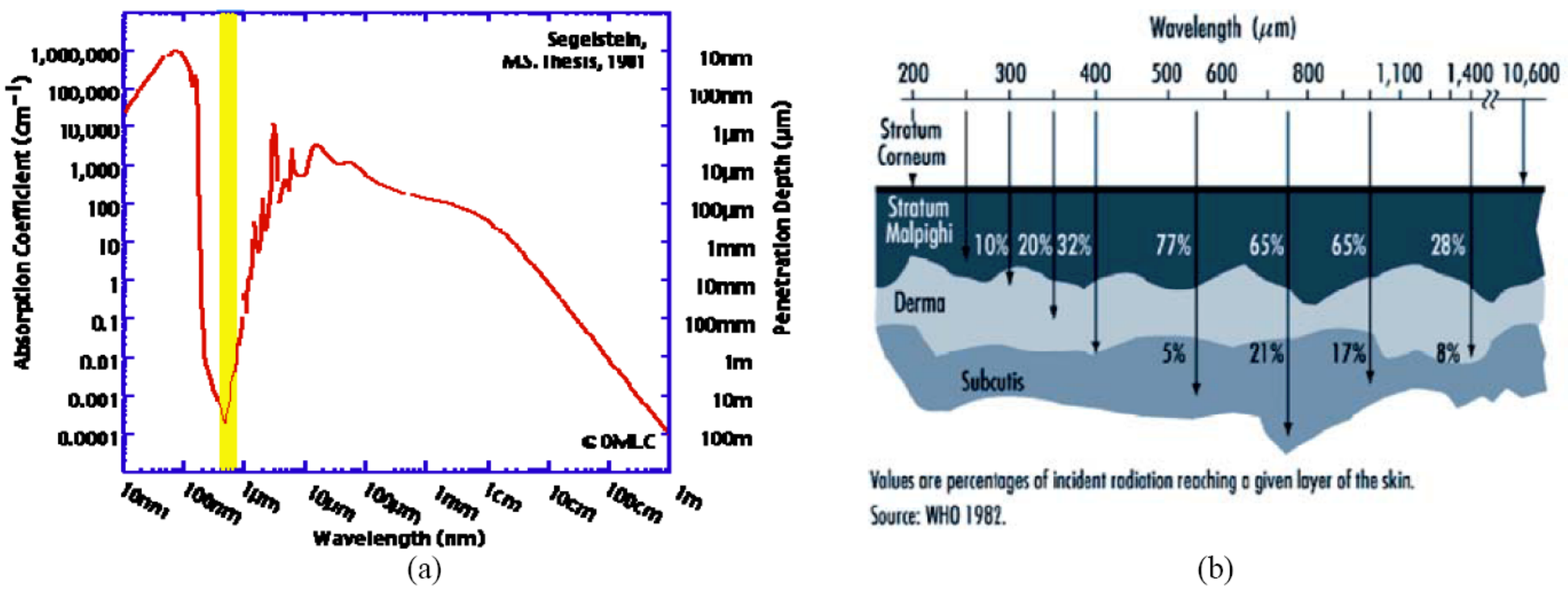}
% where an .eps filename suffix will be assumed under latex,
% and a .pdf suffix will be assumed for pdflatex; or what has been declared
% via \DeclareGraphicsExtensions.
\caption{ (a) The absorption coefficient and penetration depth of water for electromagnetic radiation of various wavelength \cite{Segelstein1981}; (b) the absorption of human skin, represented by the percentages of incident radiation with various wavelengths reaching a given layer \cite{WHO1982}.}
\label{cell_fig2}
\end{figure*}

\textbf{\emph{Choice of Frequencies:}}
We know that the water, as the major component of human body, has a very strong absorption to light with wavelength of 1,200 nm longer and 500 nm shorter \cite{Segelstein1981}. Fig. \ref{cell_fig2} (a) shows that the water absorbs 90\% of the energy of light with wavelength of 1,200 nm when travelling 1 cm. Similarly, the human skin also has a strong absorption: only 17\% of incident light with wavelength of 200 $\mu$m can reach the subcutis layer, as shown in Fig. \ref{cell_fig2} (b) \cite{WHO1982}. Further studies showed that biological systems are highly transparent to near-infrared (NIR) light with wavelength 700 nm to 1,100 nm, which can penetrate relatively deep into biological soft tissues \cite{Kam2005,Tsai2001}. Exactly, gold nanoparticles and SWCNTs have also a strong absorption peak in this region \cite{Maltzahn2009,Huang2006,Hirsch2003,Kam2005,Chakravarty2008,Zhou2009}. They also exhibit a strong fluorescent signal in this spectral region, with high signal-to-background ratio \cite{Huang2006,Iverson2013}. Therefore, the appropriate wavelength should be between 700 nm and 1100 nm. By balancing the light absorption and the efficiency of the optical antenna, we choose the wavelength of the pulsed laser to be around 900 nm. Besides, although the millimeter-wave ($\sim$100 GHz) appears another option considering the wavelength scaling effect and the advent of the SiGe heterojunction bipolar transistor technology, the difficulty in integrating the millimeter-wave source makes it infeasible for the cell-deployed S/As application.

\textbf{\emph{Layered Peer-to-Peer Network:}}
The network of S/As is operated in a layered peer-to-peer (P2P) mode. Its first layer is a peer-to-peer network, composed of the S/As responsible for information collecting, communicating and controlling/actuating. Its second layer is composed of the simple actuators for executing the single actions, and the simple sensors for converting the biological or chemical signal into the appropriate electric or optical signal. The second-layer nodes only communicate with a adjacent first-layer node. In contrast, the traditional techniques are accordant with the access-point network for the \emph{in vivo} sensing/actuating application, where each of their \emph{in vivo} nodes need to communicate with an \emph{ex vivo} node (without size limitation) acting as the access-point, so that a complex routing protocol is required. For example, the RFID system uses a large \emph{ex vivo} reader to extract the information from the \emph{in vivo} tags and, even more, to control them \cite{EPC2008}. However, the layered P2P network allows the communication to be established only between a subject sensor and another S/As within a localized volume around the subject, and it prefers the power of signal to be attenuated fast above the local distance. Thus, the loss of signal power would not be a serious problem for signal reception.

\textbf{\emph{Ultrashort Pulses for Signal:}}
Using the ultrashort laser pulses as the carrier of signal can increase the peak intensities. This helps not only to enhance the sensitivity of photodetection, but also to reduce the thermal effect due to absorption. In addition, we dichotomously code the pulses by "0" and "1" according to whether they have a decent amount of power, so as to bear a unity signal-to-noise ratio.

\textbf{\emph{Radiation Pattern Switching:}}
The optical antenna arrays for signal transmission can realize different radiation patterns switched by controlling the static voltages applied on the antenna arrays (due to their nonlinear optical characteristics). Accordingly, by switching the radiation pattern, a transmitter is able to focus the transmitted power on the specific directions and hence optimized the power of signal received by its recipient. If all the recipients and their optimal radiation patterns are related and memorized by the transmitter, the signal transmission becomes smart and more directional. As a result, the received signal power would be enhanced and make up the absorption loss.

%===========================================================================================================
\section{Considerations of Cross-Layer Designs}

Fast response rate of the actuator is usually needed when the sensor transmits the command. In the instance of excitation suppression, we want the actuator to respond before the action potential arrives, so that the response time should be shorter than the traveling time of the action potential from the sensing spot to the actuation spot. The response rate is associated with the signal collision, communication rate, signal complexity, etc. In particular, the communication system should handle the dense collisions of signal, because the activities that excite the actions of nearby sensors may be, to some extent, interrelated. For example, an action potential travelling along a nerve fiber may excite a series of nearby sensors (mounted with the nerve cells around) within a very short time interval.

\textbf{\emph{Layered Peer-to-Peer Network:}}
The layered P2P network can greatly improve actuation speed, compared to the access-point network. First, there is high probability of signal collisions, if all S/As have to communicate with an access point (or base station) and only one stable channel of communication can be provided (because the frequency resource is greatly limited by the microscale and \emph{in vivo} optoelectronic device). However, if the sensors directly communicate with their target actuators and the signal powers are much localized, the signal collisions can merely happen among S/As in a small volume and the number of collisions would be largely reduced. The collisions, like the traffic jams, result in the time delay of communication, as they should be overcome by the time-division strategy (i.e., to retransmit after proper delays when encountering collisions, or to avoid the collision by scheduled rolling of channel access). Thus, the layered P2P network can enhance the actuation speed by parallel communications with each actuator as a subcenter. Second, the layered P2P network can avoid the time spent on signal forwarding.

\textbf{\emph{Simplification of Communication Protocols:}}
The shorter messages and the simpler verification process brings faster actuation. This has been discussed above.

\textbf{\emph{Ultrafast Laser:}}
Generally, the optical frequency allows high communication rate. We use the ultrafast pulsed laser with the pulse width in the order of 100 fs. Accordingly, the repetition rate of pulses, also as the bit rate of communication, can easily surpass 1 MHz.  Thus, the command from the sensor to the actuator can be transmitted in a short time interval. Nevertheless, the main restraint is the response rate of photodetectors without voltage bias.

\textbf{\emph{Radiation Pattern Switching:}}
From the perspective of network topology, this method can effectively reduce the nodes connected to the subject node (their transmitted power can be detected by the subject node) and hence reduce the probability of signal collisions.

\textbf{\emph{Partial Space-Division:}}
We realize a partial space-division of communication channels by using multiple photodetectors in a proper arrangement for signal reception. For example, we can arrange two photodiodes on the top and bottom of the chip, so that the signals received by the two sides can be differentiated by comparing the two photocurrents. Thus, the signal collisions can be detected and even recovered in many cases, especially when the incident signals are transmitted to the distinct sides of the receiver. Nevertheless, the criteria of `0' and `1' for each photodetector should be carefully studied due to the optical scattering in the biological environment, and for better detection, this method should be properly assisted with other collision-detection methods such as the signal-format verification.

\textbf{\emph{Learning Mode of S/As:}}
In order to know the physical position and nearby topology of a S/A, we design a learning mode operated before they can normally work. In this mode, the S/As can learn about their physical position, the local network topology, the working submodes, and the radiation pattern optimized for each recipient, etc. The `user interface' plays an important role in the learning mode, such as to scan positions, to excite the mode and to output the learning result.

\textbf{\emph{Collision Detection when Muted and Exit when Colliding (CDWM/EWC):}}
To coordinate with our dichotomous coding method (`0' ¨C without power, `1' ¨C with power), we propose a novel method for collision management, in which each transmitter detects the incident power when `0' is transmitted. Because the collision of multiple `0's or multiple `1's would not influence the common recipient, this approach can ideally detect the collisions of `0' and `1', and implement signal nonblocking transmission if all other transmitters are visible (i.e., no hidden node exists).

\textbf{\emph{Control Approaches to Signal Collisions Caused by Hidden nodes:}}
We should consider more solutions for the signal collision due to the existence of the hidden nodes (especially when the method of radiation pattern switching takes effects). Aside from the commonly used `handshake' protocol and the signal-format verification for reception, we invent the notifying/blocking signal, similar to the RTS/CTS (Request to Send and Clear to Send). An S/A transmits a blocking signal to all the nearby S/As as soon as it receives a notifying signal; the blocking signal then blocks signal transmission of all the nearby S/As except the one sending the notifying signal. However, different from the RTS/CTS, we consider the collisions of notifying signals (or RTS) due to the dense collisions and optimize the processing sequence for higher speed.

In the following parts, we elaborate the basic principle of the sensor/actuator, the optical antenna, and the communication protocol, which are all optimized for this \emph{in vivo} sensing/actuating application. Among all of the components, the optical antenna is the core of our proposed solution and its performance determines the success of the system.

%===========================================================================================================
%                                The Second Section (Sensor/Actuator)
%===========================================================================================================
\section{Sensors/Actuators}

%\subsection{Subsection Heading Here}
%Subsection text here.

%\subsubsection{Subsubsection Heading Here}
%Subsubsection text here.

\begin{figure*}[!t]
\centering
\includegraphics[width=14.0cm]{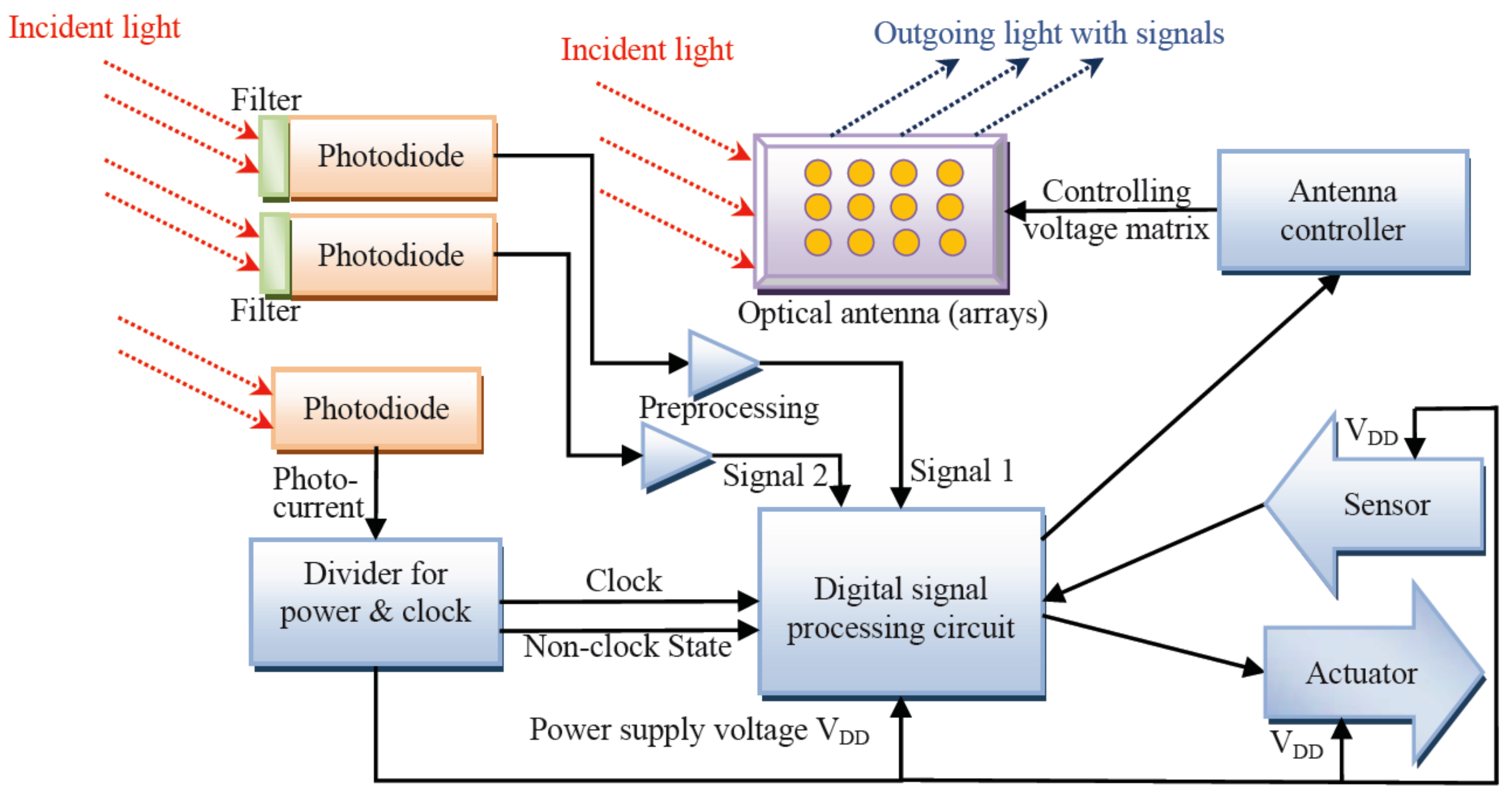}
% where an .eps filename suffix will be assumed under latex,
% and a .pdf suffix will be assumed for pdflatex; or what has been declared
% via \DeclareGraphicsExtensions.
\caption{Block diagram of a sensor/actuator.}
\label{cell_fig3}
\end{figure*}

\begin{figure*}[!t]
\centering
\includegraphics[width=14.0cm]{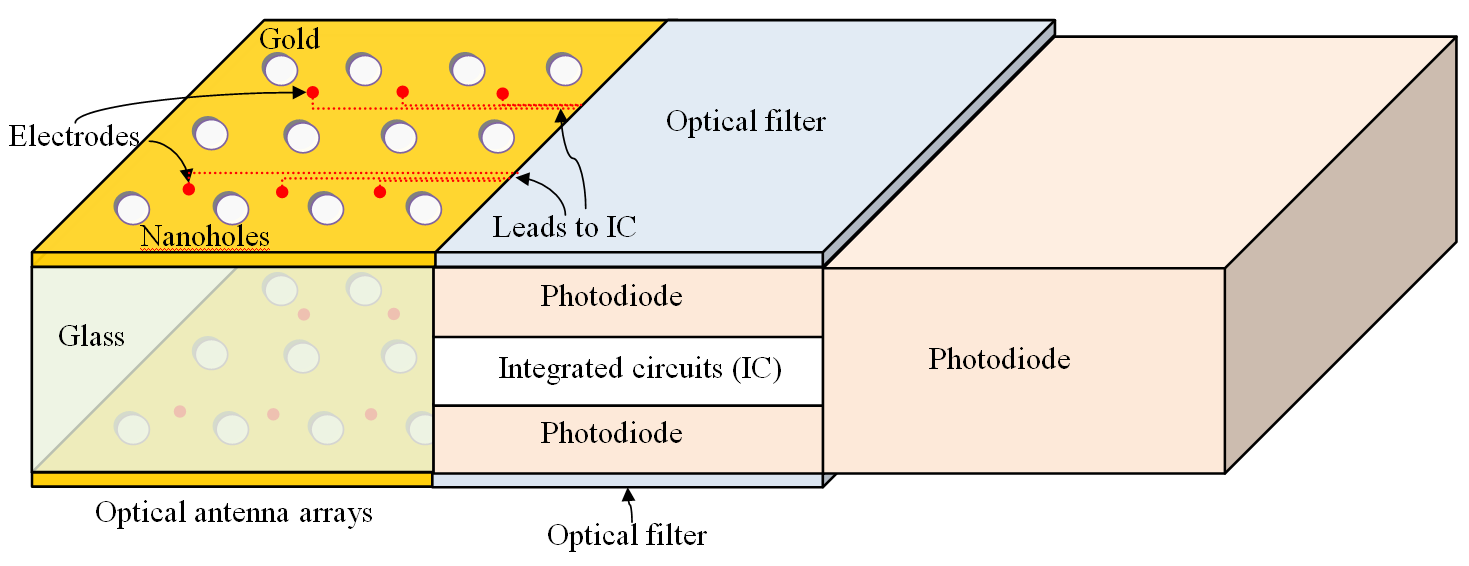}
% where an .eps filename suffix will be assumed under latex,
% and a .pdf suffix will be assumed for pdflatex; or what has been declared
% via \DeclareGraphicsExtensions.
\caption{Physical structure of a sensor/actuator.}
\label{cell_fig4}
\end{figure*}

In our design, a first-layer S/As should work as shown in Fig. \ref{cell_fig3}, and its physical structure could be like Fig. \ref{cell_fig4}. The incident laser beam (a continuous sequence of pulses) is received and converted into electricity by a photodiode (shown on the left bottom of Fig. \ref{cell_fig3} and on the right side of Fig. \ref{cell_fig4}), and the circuit of the `divider for power and clock' then divides this electrical signal into a power-supply DC voltage, the clock signal, and the signal of non-clock state (due to a long absence of the clock signal).

In addition, the S/A communication signals are carried by a specific frequency of light pulses transmitted by the optical antenna (arrays), but the intensities of these signals are much lower than the incident laser pulse. Thus, we use the optical filters to select the desirable frequency so that they can be well received by the photodiode (shown on the left top of Fig. \ref{cell_fig3} and on the middle of Fig. \ref{cell_fig4}). The optical filters could be a sort of opaque dielectric which has characteristic absorption covering the incident frequency. We arrange two photodiodes on both top and bottom of the chip (see Fig. \ref{cell_fig4}), so that the optical signals received by the two sides can be differentiated by comparing the two photocurrents. Therefore, the collision of signals can be possibly detected and even recovered, thus realizing a partial space-division of communication channels.

\begin{figure*}[!t]
\centering
\includegraphics[width=14.0cm]{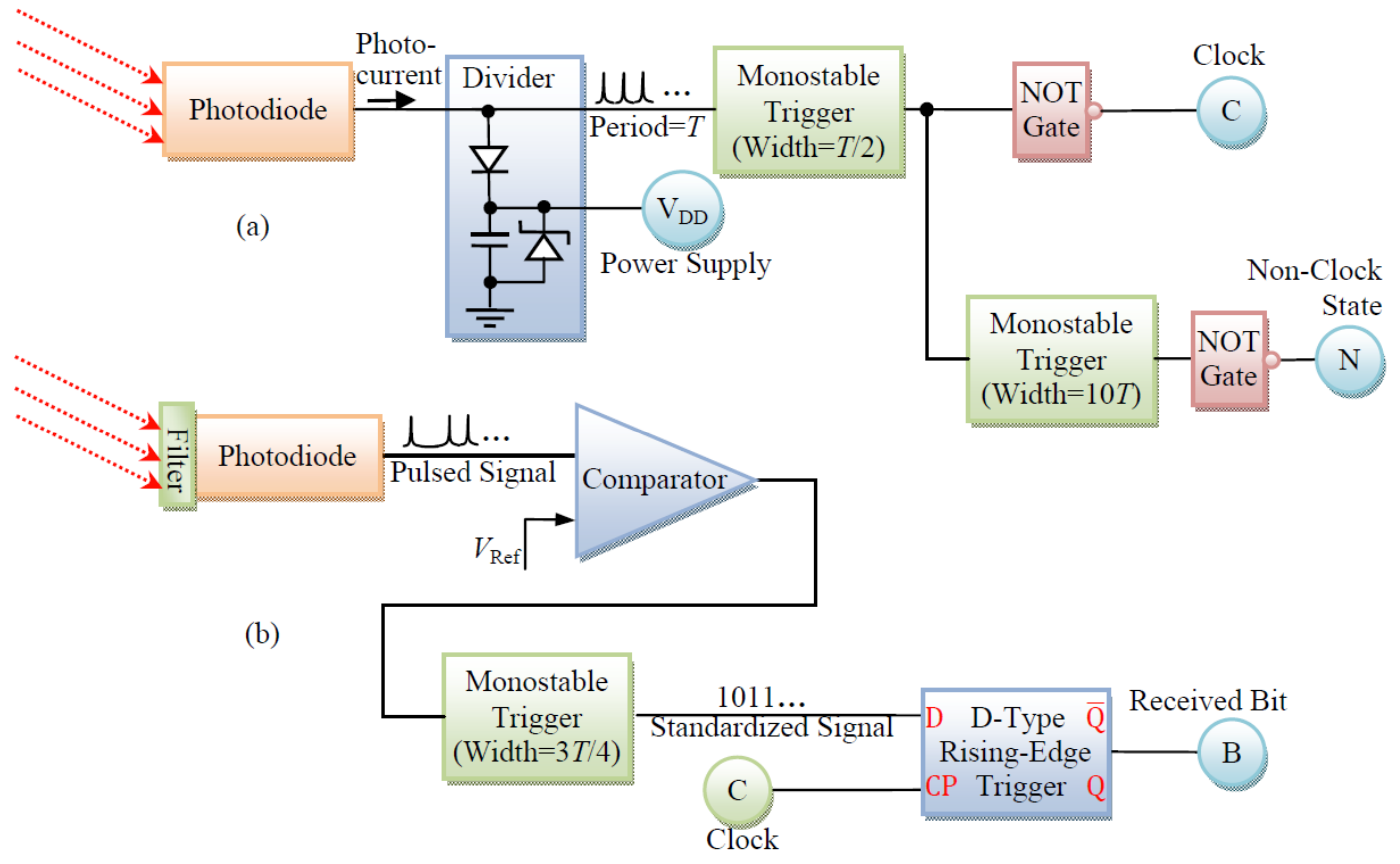}
% where an .eps filename suffix will be assumed under latex,
% and a .pdf suffix will be assumed for pdflatex; or what has been declared
% via \DeclareGraphicsExtensions.
\caption{Block diagram of the preprocessing circuit for (a) power and clock (b) communication signals.}
\label{cell_fig5}
\end{figure*}

The photocurrents obtained from the photodiodes are all inputted into the `digital signal processing circuit', but some proper preprocessing should be executed beforehand (see Fig. \ref{cell_fig5}). On the one hand, the power received by the photocurrent can be divided into the power-supply DC voltage and the clock signal, by a simple combination of a capacitor and two diodes. A voltage buffer can be further added to the DC voltage output for better stabilization of voltage, while it may consume more power. Then, through a monostable trigger combined with a NOT gate, the ultrashort-pulse sequence is converted to the clock of the digital circuit. Its rising edge happens at half a cycle after the laser pulse, so as to allow sufficient delay in the signal transmission (through the optical antenna arrays) at the advent of the next laser pulse. A second monostable trigger can detect a long absence of clock signal (as a non-clock state) which is used to synchronize the clocks of all S/As or to excite the learning mode.

On the other hand, the communication signal is compared to a reference voltage after the photoelectric conversion, in order to cancel the influence of optical scattering by biological substances; otherwise, the light incident from the top and bottom can be hardly differentiated by the two photodiodes. The following monostable trigger standardizes the signal with the rising edge happening one quarter cycle after that of the clock and one quarter cycle before the next laser pulse, thus reserving sufficient time margin for signal processing. The digital signal is then latched by the D rising-edge trigger until the rising edge of the next clock. If the trigger is replaced by a shift trigger, the bit data can be stored for more cycles and the inputted data could be read in the unit of word.

Furthermore, the `digital signal processing circuit' is responsible for deciding when and how to respond to the sensed information, and to communicate with other S/As, and to execute the proper actuation. The circuit design should depend on what communication protocol is adopted and how the sensor part or the actuator part works. In order to minimize the size to accommodate with cell, we should optimally simplify the communication protocol. Nevertheless, the design should follow the conventional method for digital integrated circuit, and consider the state-of-art silicon technology (e.g., 100 nm). The device size can be limited within 100 square micron, a cell's scale, if less than 1000 transistors are used.

Finally, the optical antenna plays a vital role in the S/A. It not only transmits the signals by an optical carrier different from the incident laser frequency and converts the signal from electricity to light in the scale of microns, but also adapt the radiation pattern to better aim at the recipient. In principle, it utilizes the nonlinear optical properties of metals and the interaction between photons and plasmons. In the following part we try to give tentative solutions, but the verification and parameter determination should rely on the physical modeling and the device-level simulation.

For the less smart second-layer S/As, we can directly adopt gold nanorod antennas \cite{Maltzahn2009,Huang2006} or SWCNT antennas \cite{Zhao2012,Chakravarty2008,Zhou2009}, with antibody-functionalized, for the cancer cell detection and photothermal ablation. We can also use ultra pH-sensitive nanoparticles with cancer-targeting as sensors for cancer cell tracking and positioning \cite{Wang2014}, and magneto-electric nanoparticles as actuators for anti-cancer drug delivery and controlled release \cite{Guduru2013}.

%===========================================================================================================
%                                The Third Section (Optical Antenna Arrays)
%===========================================================================================================
\section{Optical Antenna Arrays}

The optical antenna first couples the optical far-field to the near-field so as to effectively interact with the plasmons in antenna, and it then transmits the near-field to the free space with intensity focused in some expected directions (i.e., radiation patterns). In other words, the optical antenna, rather than convert electrical signals into the form of electromagnetic wave as the RF antenna, converts light (electromagnetic wave) to light but with different frequencies and directions (which are electrically controllable). Further, we compose the optical antenna of nanoantenna arrays (e.g., holes in gold plane or gold particles) which can be controlled separately by the applied voltages, so that the optical antenna can convert the electrical signal into the optical forms without directly controlling the light source.

Besides, we also hope that the optical antenna is a ¡®smart antenna¡¯ that can intelligently adapt the radiation pattern toward the recipient. The RF ¡®smart antennas¡¯ modulates the signal phase for each array element in order to obtain various radiation patterns, but imitation is infeasible since the period of optical radiation is too short to be controlled. However, we design a novel approach which utilizes antenna arrays and the nonlinear optical properties of metal materials such as gold. The radiation patterns and other properties of the optical antenna can be changed by varying the quasi-static electric field applied on each antenna array.

In the nonlinear optical materials, the induced polarization depends linearly on the electric field strength in a manner that can often be described by the relationship

\begin{equation}\label{cell_eq_1}
P_{I} = \!\!\!\sum_{IJKL}\!\!\epsilon_0(\chi_{IJ}^{(1)}E_J+\chi_{IJK}^{(1)}E_J E_K+\chi_{IJKL}^{(1)}E_J E_K E_L+\cdots \nonumber
\end{equation}

where subscripts $I$, $J$, $K$, $L$ represent any of the coordinate index $X$, $Y$, $Z$, $\chi_{IJ}^{(1)}$ is the linear susceptibility, and $\chi_{IJK}^{(1)}$ and $\chi_{IJKL}^{(1)}$ are second- and third-order nonlinear susceptibilities respectively. The polarization is closely related with the intensity of interaction between the optical field and the material, although it is usually otherwise described in terms of the permittivity which is the ratio of the polarization and the electric field (but the permittivity is a tensor in this case). It is obvious that there are more frequency components than the incident optical field in the formulation of the polarization. In particular, the outgoing scattered optical field can have a different frequency from the incident field and it can be effectively changed by applying a strong static field. In addition, the real part influences both the resonating wavelength and the radiation efficiency, while the imaginary part of the polarization is associated with the absorption of incident field. Because the scattered fields from all antenna arrays are coherent, the radiation pattern can be changed by adjusting the geometrical distribution of the array elements, by adjusting the radiation pattern and efficiency of a single array element (through changing its geometry and the laser frequency), and by varying the nonlinear absorption of the array elements (through applying the static potentials).

The third-order nonlinear susceptibility of gold ($\chi^{(3)}\sim 1$nm$^2/$V$^2$) \cite{Palomba2011}, for instance, is more than 3 orders of magnitude larger than that of the many nonlinear optical crystals, such as lithium niobate (LiNbO$_3$) \cite{Ganeev2003}. Laser systems employ nonlinear crystals instead of noble metals for frequency conversion because they are transparent, enabling them to be placed in a beam line, and because they allow phase matching¡ªthe coherent addition of the nonlinear response on propagation through a periodic crystal. For phase matching to occur, the crystal needs to be many wavelengths in size. In plasmonics, one is usually more concerned with local nonlinear signals, and without the constraints associated with lasers, exploiting the nonlinearities of noble metal antenna structures may be favorable \cite{Bharadwaj2009}. Previous investigation on the nonlinear optical antenna focuses on the physical phenomena including second-harmonic generation, third-harmonic generation, two-photon excited luminescence, and four-wave mixing \cite{Palomba2011,Bharadwaj2009,Palomba2009}, rather than  its interaction with the static electric field and considering it as an optoelectric device to engineer the scattering and absorption of incident light. Thus, it is a new land to utilize the nonlinear optical properties of the optical antenna arrays for coding signals and controlling radiation patterns by applying the electrical potential matrix.

\begin{figure}[!t]
\centering
\includegraphics[width=8.9cm]{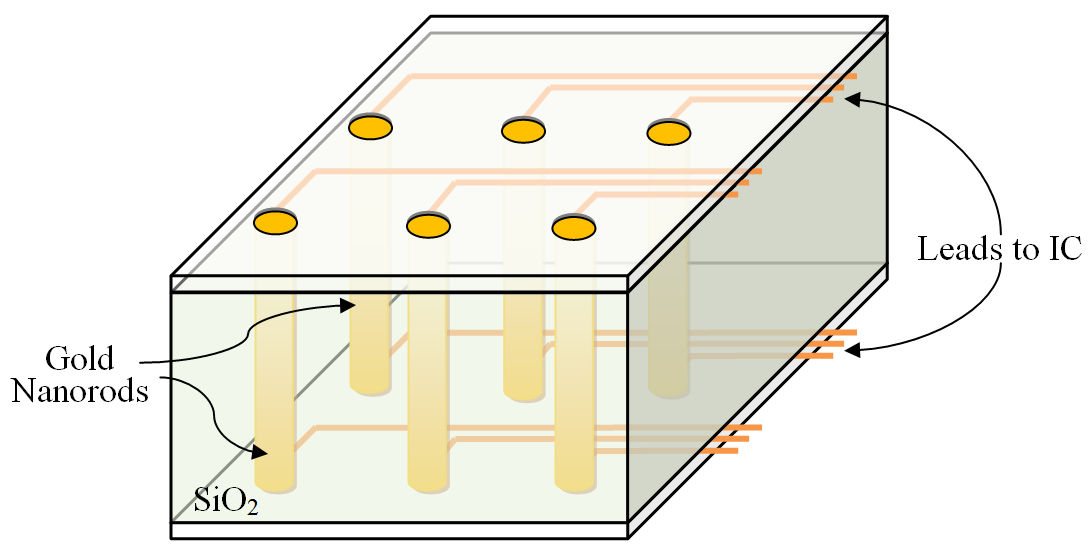}
% where an .eps filename suffix will be assumed under latex,
% and a .pdf suffix will be assumed for pdflatex; or what has been declared
% via \DeclareGraphicsExtensions.
\caption{Another design for the optical antenna arrays composed by gold nanorods in SiO$_2$}
\label{cell_fig6}
\end{figure}

In order to carry on the further research, we need the physical modeling of the optical antenna with a proper accuracy. Then, we need to carefully engineer the geometry of the antenna arrays, according to our design principles discussed above. One of the possible designs is shown in Fig. \ref{cell_fig4}. The incident laser beam is obliquely projected onto the gold plane with arrays of nanoholes (on one side), and the photons effectively interact with the plasmons in the gold. A nanohole can be seen as an antenna array element, analogy to the circular antenna. This geometry makes the energy of the induced near-field mainly stored in the magnetic field, so that its interaction with the biological environment is weak. The induced optical far-field (i.e., the scattered field) has multiple frequency components, but only one can resonate with the nanohole (whose size is usually 2$\sim$6 times shorter than the free space wavelength \cite{Novotny2007}). One part of the field is backscattered and another part penetrates the glass below the gold plane and resonates again with the gold plane on the other side so that it can be radiated into the free space. In other words, the optical antenna arrays composed by the gold nanoholes can select the characteristic wavelength from the multiple-frequency components induced by the nonlinear effect. Besides, by applying a matrix of electric potentials on the antenna arrays, the radiation can be concentrated in specific directions or be coded as `1' or `0'. From the practical perspective, this antenna arrays are simple to realize, while the design of electrodes becomes challenging. Moreover, Fig. \ref{cell_fig6} shows another possible design, which adopts the gold nanorods and applies voltages on the both ends of each rod. The wavelength of radiation tend to seek resonances with the length of the nanorod rather than its diameter. The fabrication may involve photolithography and laser drilling \cite{Yu2009}. Nevertheless, before the real fabrication, we need to construct a physical model with proper accuracy, and use it to verify the design and to engineer the related parameters, including the laser frequency, the antenna dimension and geometry, the arrangement of arrays, and the position of electrodes.

Modeling the optical antenna composed of the nanoscale metal elements with strong nonlinear optical effects is quite different from the modeling the nonlinear optical crystals, because the oscillation of electrons (plasmons) is induced and interacts with both phonons and photons in a nonlinear manner (i.e., with frequency-talk). Nevertheless, it could follow our previous theoretical work on the linear optical antenna composed of a carbon nanotube \cite{Zhao2012}, but the consideration on the nonlinear susceptibilities and the photon excitation should be added. Thus, a semiclassical model should be used for the balance between accuracy and computational complexity. On the one hand, the optical field can be treated as the classical electromagnetic wave. More accurately, the ultrashort laser pulse is modeled as the ultrashort wave packet. The reason is that the wave nature of the optical field for the metal nanoparticles is much more important than its quantum properties, since the optical field tends to seek resonances with the antenna geometry rather than to directly excite photocurrent like in semiconductors (the excitation and absorption of photons can be merged into the scattering terms in the transport equation). On the other hand, the plasmonic transport in the nanoantenna should be considered using the Wigner function method \cite{Querlioz2010} or the Liouville equation \cite{Grubin1993}, due to the notable interband transitions. Up to the third-order susceptibilities should be considered. Still, it is possible that these complex methods can be simplified by the Boltzman equation or even a quasi-Ohm's relation in the extent of allowable accuracy \cite{Zhao2012}, and this is what we need to examine.

%===========================================================================================================
%                                The Fourth Section (Communication Protocol)
%===========================================================================================================
\section{Communication Protocol}

\begin{figure}[!t]
\centering
\includegraphics[width=8.9cm]{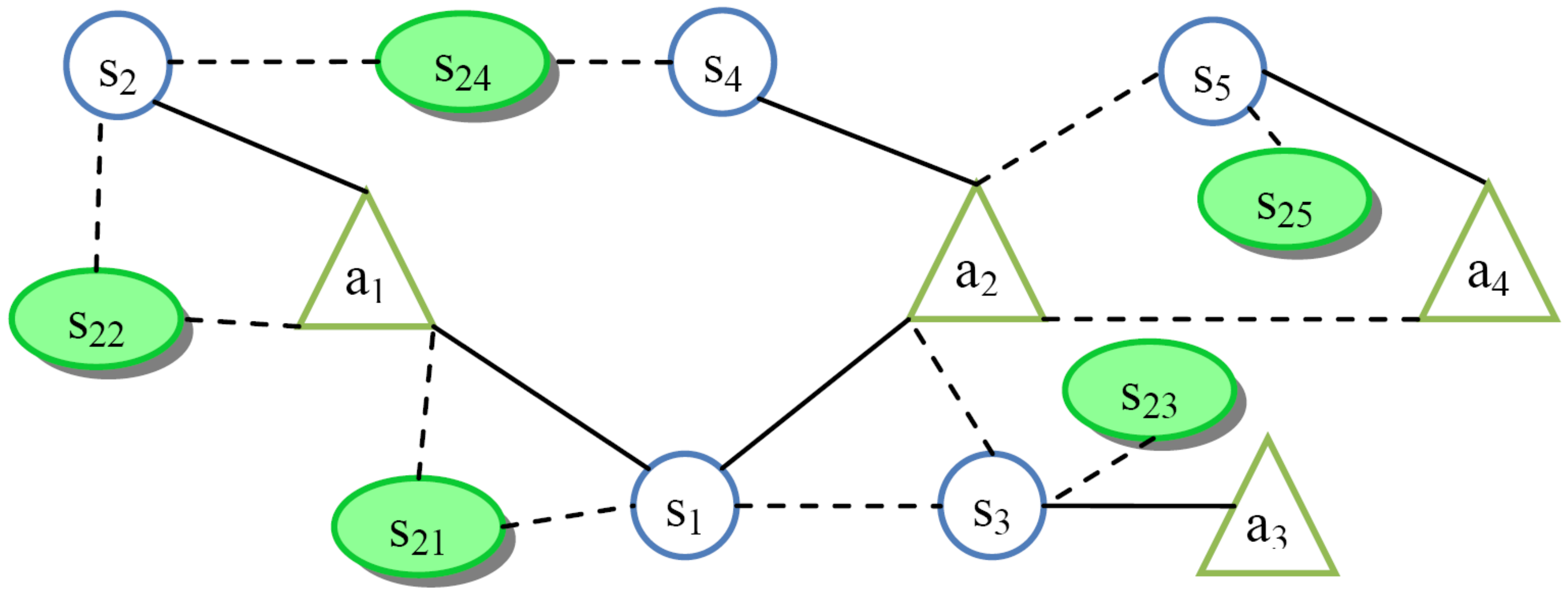}
% where an .eps filename suffix will be assumed under latex,
% and a .pdf suffix will be assumed for pdflatex; or what has been declared
% via \DeclareGraphicsExtensions.
\caption{An example of the topology of the sensor/actuator (S/A) network. The sensors and actuators (denoted by circles and triangles, respectively) are coded by the combination of an alphabet (`s' or `a') and a number. The solid-line connector between a sensor and an actuator indicates the actuator is a recognized recipient for the sensor, whereas the dashed-line connector between a pair of S/As indicates their communication is unrecognized even though signals can be physically transmitted through.}
\label{cell_fig7}
\end{figure}

In order to design a proper communication protocol for the S/As, we need to better understand the possible topology of the network. Although the S/As are described as equivalent in Fig. \ref{cell_fig1}, they should be clearly differentiated in the communications due to their different tasks. A more detailed topology is presented by the example in Fig. \ref{cell_fig7}, where s$_1 \sim$ s$_5$ are first-layer sensors, a$_1 \sim$ a$_4$ are first-layer actuators, s$_{21} \sim$ s$_{25}$ are second-layer sensor/actuator clusters, a first-layer sensor may control multiple actuators, and an actuator may be controlled by multiple first-layer sensors (shown as solid line connector). Meanwhile, an S/A may be influenced by the signal of a non-recognized recipient (shown as dashed line connector). We have designed a learning mode of S/As, which is operated before the S/As work normally. In this mode, the S/As can learn about the local network topology, the working mode, the radiation pattern optimized for each recipient, etc. The `user interface' plays an important role in the learning mode. It can transmits signals to S/As by coding a second laser beam and receives the signal by using the \emph{ex vivo} photodetectors, respectively for the input and the output of data. The switching of modes can be notified by a long absence of the clock signal.

In the working mode, we need a novel communication protocol for the particular application, in order to arrange the limited communication resources appropriately and concert the working of related S/As, and especially to solve collisions of signal in the common channel. It should combine our basic strategy of the time-division and the partial space-division. We have designed two different protocols -- a synchronous one and an asynchronous one, depending on whether the signal is frame synchronous for transmission and reception or not. As for a more practical design, we should consider protocols with extended functions, such as the message relay to the `user interface' from the nodes lacking a direct connection.

\subsection{Learning Mode}

In order to obtain the learning result (stored in the static memory of S/As) as the examples given in TABLE I $\sim$ III (for the network in Fig. \ref{cell_fig7}), we have designed the following steps of the learning mode.

\begin{table*}[!htbp]
\centering
\caption{Information stored for the sensor {\rm s$_1$}.}
\label{tab-1}
\begin{tabular}{cccccccccc}
\toprule
S/R Code & s$_1$ & s$_2$ & s$_3$ & s$_4$ & s$_5$ & a$_1$ & a$_2$ & a$_3$ & a$_4$\\
\midrule
\rowcolor{mygray}
Binary Address & 0000 & 0001 & 0010 & 0011 & 0100 & 1000 & 1001 & 1010 & 1011\\
Recognized Recipient?&0&0&0&0&0&1&1&0&0\\
Physical Recipient?&0&0&1&0&0&1&1&0&0\\
Optimal Pattern&0&0&1&0&0&3&2&0&0\\
\bottomrule
\end{tabular}
\end{table*}

\begin{table*}[!htbp]
\caption{Information stored for the actuator {\rm a$_1$}.}
\label{tab-2}
\centering
\begin{tabular}{@{}cccccccccc}
\toprule
S/R Code & s$_1$ & s$_2$ & s$_3$ & s$_4$ & s$_5$ & a$_1$ & a$_2$ & a$_3$ & a$_4$\\
\midrule
Binary Address & 0000 & 0001 & 0010 & 0011 & 0100 & 1000 & 1001 & 1010 & 1011\\
Recognized Recipient? &1&1&0&0&0&0&0&0&0\\
Physical Recipient? &1&1&0&0&0&1&0&0&0\\
Optimal Pattern &2&1&0&0&0&0&3&0&0\\
\bottomrule
\end{tabular}
\end{table*}

\begin{table*}[!htbp]
\caption{working modes and physical positions of first=layer nodes.}
\label{tab-3}
\centering
\begin{tabular}{@{}cccccccccc}
\toprule
S/R Code & s$_1$ & s$_2$ & s$_3$ & s$_4$ & s$_5$ & a$_1$ & a$_2$ & a$_3$ & a$_4$ \\
\midrule
Binary Address & 0000 & 0001 & 0010 & 0011 & 0100 & 1000 & 1001 & 1010 & 1011 \\
Working Mode & T1 & T2 & T3 & T3 & T1 & T3 & T2 & T1 & T3 \\
Physical Position & 3 & 14 & 4 & 16 & 18 & 8 & 11 & 5 & 13 \\
\bottomrule
\end{tabular}
\end{table*}

%% An example of a floating table. Note that, for IEEE style tables, the
%% \caption command should come BEFORE the table. Table text will default to
%% \footnotesize as IEEE normally uses this smaller font for tables.
%% The \label must come after \caption as always.
%\begin{table}[!t]
%% increase table row spacing, adjust to taste
%\renewcommand{\arraystretch}{1.3}
%% if using array.sty, it might be a good idea to tweak the value of
%% \extrarowheight as needed to properly center the text within the cells
%\caption{An Example of a Table}
%\label{table_example}
%\centering
%% Some packages, such as MDW tools, offer better commands for making tables
%% than the plain LaTeX2e tabular which is used here.
%\begin{tabular}{|c||c|}
%\hline
%One & Two\\
%\hline
%Three & Four\\
%\hline
%\end{tabular}
%\end{table}

\textbf{\emph{Physical Position Learning:}}
To positioning S/As, the therapeutic body area is divided into honey grid, whose cell size is determined by laser spot size. The signal laser of `user interface' scans the grid cell one by one and notice S/As in every grid cell their physical position number. There position information may be used, for example, for increasing irradiation power to specific position so to destroy tumour cells, based on the sensor's request.

\textbf{\emph{Local Topology Learning:}}
S/As should automatically learn which S/As are physically connected to them. This can be done by a special procedure of communication: a subject S/A sequentially transmits all the addresses except for itself, and an acknowledge signal (ACK) would be transmitted back if their addresses are received by other S/As; the subject S/A regards the address as a physical recipient after the ACK is received; after completion of all addresses, the learning subject is changed to another S/A. Thus, the local topology of each S/A can be probed. It is notable that we only record the `worst' topology for all the radiation patterns so that each S/A has the most connectors (also in Table 1 and 2), because it costs too much to store and search all topologies.

\textbf{\emph{Direction Learning:}}
The optimized radiation pattern for each recipient can be learnt from a sequential trial of all patterns. The signal intensities for different patterns are then compared by the recipient and the learning subject can obtain the pattern number from the feedback sent by the recipient.

\textbf{\emph{Working Mode Learning:}}
In the procedure of physical position learning, local topology learning, direction learning, the working mode of every first-layer S/A can be configured to reach a appropriate time-space diversity and reduce collision probability.

\subsection{Working Mode}

In order to simplify as much as possible the complexity of the devices and make full use of the different operating environments of the first-layer nodes and the second-layer nodes, we designed a synchronous communication protocol to work in the synchronous mode and an asynchronous communication protocol to work in the asynchronous mode. The synchronous protocol focuses on making a reliable yet simple time division multiaccess on the common communication channel among the first-layer nodes. The asynchronous protocol match to the less smart second-layer nodes, where full frame synchronization is an unbearable burden.

\begin{figure*}[!t]
\centering
\includegraphics[width=10.0cm]{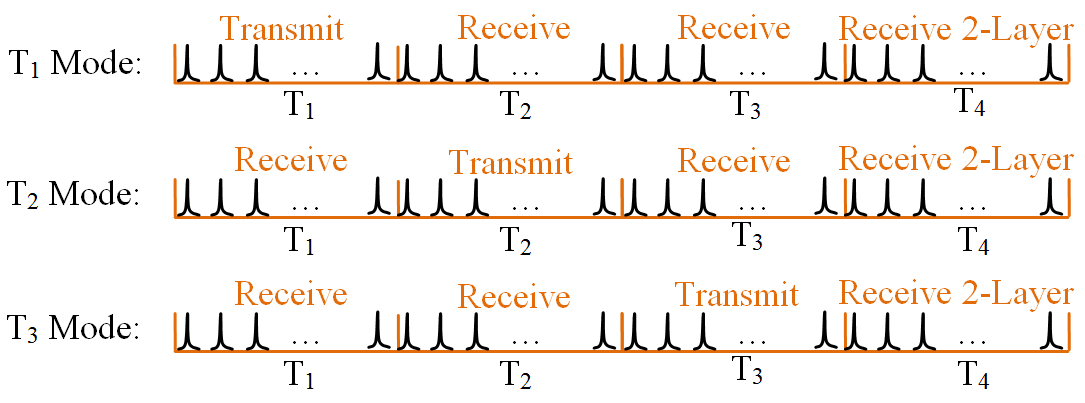}
% where an .eps filename suffix will be assumed under latex,
% and a .pdf suffix will be assumed for pdflatex; or what has been declared
% via \DeclareGraphicsExtensions.
\caption{Instruction cycles for a sensor/actuator with T$_1$, T$_2$, or T$_3$ Mode, for which to transmit and to receive signals are complementary in subcycle T$_1$, T$_2$, and T$_3$. In order to receive the second-layer signal, T$_4$ are reserved. Each of the four parts T$_1$ $\sim$ T$_4$ has several clock cycles depending on the maximum length of instructions.}
\label{cell_fig8}
\end{figure*}

\begin{figure*}[!t]
\centering
\includegraphics[width=10.0cm]{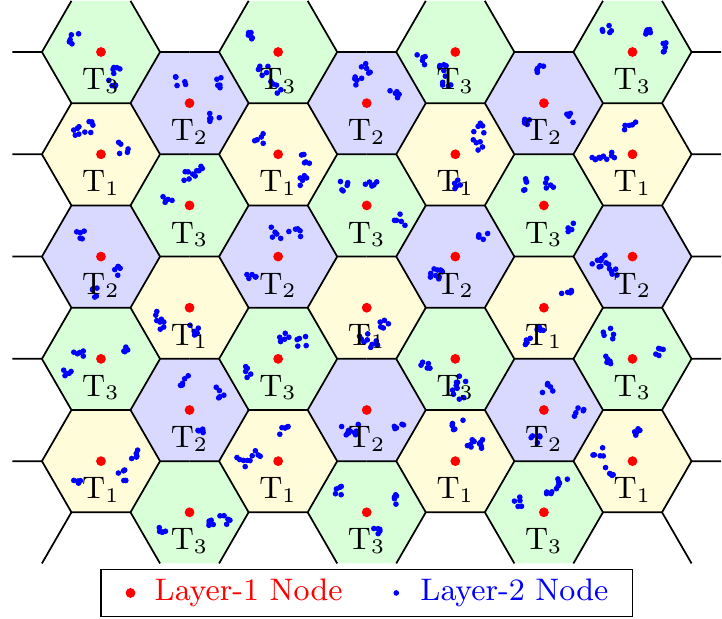}
% where an .eps filename suffix will be assumed under latex,
% and a .pdf suffix will be assumed for pdflatex; or what has been declared
% via \DeclareGraphicsExtensions.
\caption{A rational time-space diversity architecture, where the different color of honey cells represents the different mode of the first-layer node in that honey cell (T$_1$, T$_2$, or T$_3$ Mode).}
\label{cell_fig9}
\end{figure*}

\textbf{\emph{Synchronous Protocol:}}
For the first-layer peer-to-peer network, all S/As share a half-duplex communication channel and a simplex actuator command channel to second-layer actuators, which use the time diversity and the spatial diversity to achieve multiple access. This approach makes all first-layer S/As synchronized and operated in the same instruction cycles. It takes the advantage that all the S/As use the same clock cycles produced by a pulsed laser and the time difference due to distance is negligible. As shown in Fig. \ref{cell_fig8}, an instruction cycle include four subcycles, T$_1$ $\sim$ T$_4$. In the learning Mode, A first-layer S/A can be configured to operate in the T$_1$, T$_2$, or T$_3$ working mode, which transmits an instruction to another S/A in T$_1$, T$_2$, or T$_3$, and receives instructions from other S/As in other three subcycles. In such layout of time-division, a S/A is transmitting signals while its recipient S/As is receiving them, and vice versa. In the next subcycle after a successful reception, the received instruction is decoded to generate the instruction into its transmitting queue (TQ), since the decoding circuit in S/As also use the same clock cycle produced by laser pulses. Combining this time diversity with a rational spatial diversity, as shown in Fig. \ref{cell_fig9}, form a maximum possible collision avoidance architecture. The T$_4$ subcycle is used for the second-layer nodes to send instructions to their first-layer subject node.

\begin{figure*}[!t]
\centering
\includegraphics[width=10.0cm]{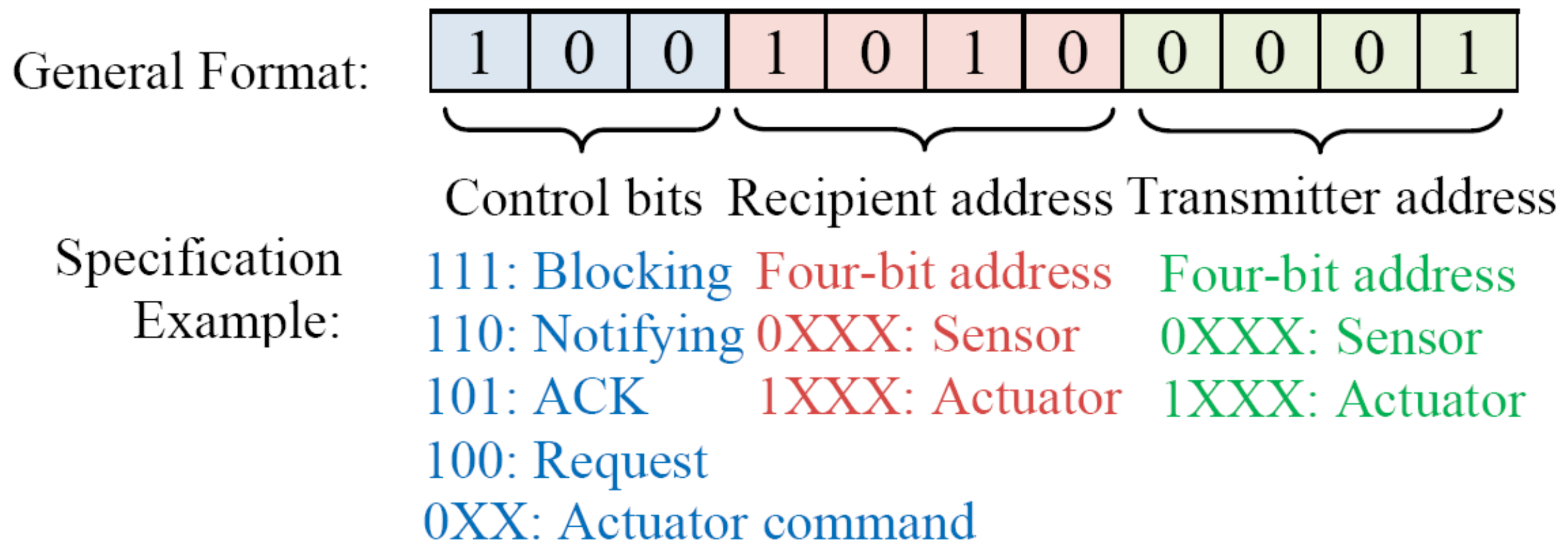}
% where an .eps filename suffix will be assumed under latex,
% and a .pdf suffix will be assumed for pdflatex; or what has been declared
% via \DeclareGraphicsExtensions.
\caption{Format of major instructions in the synchronous protocol.}
\label{cell_fig10}
\end{figure*}

The format of instructions for transmission is shown in Fig. \ref{cell_fig10}. The instruction includes three parts: recipient address, control bits, and transmitter address. The recipient address is placed at the beginning because we hope the recipient to fast decide whether the instruction is for it.

\begin{figure*}[!t]
\centering
\includegraphics[width=16.0cm]{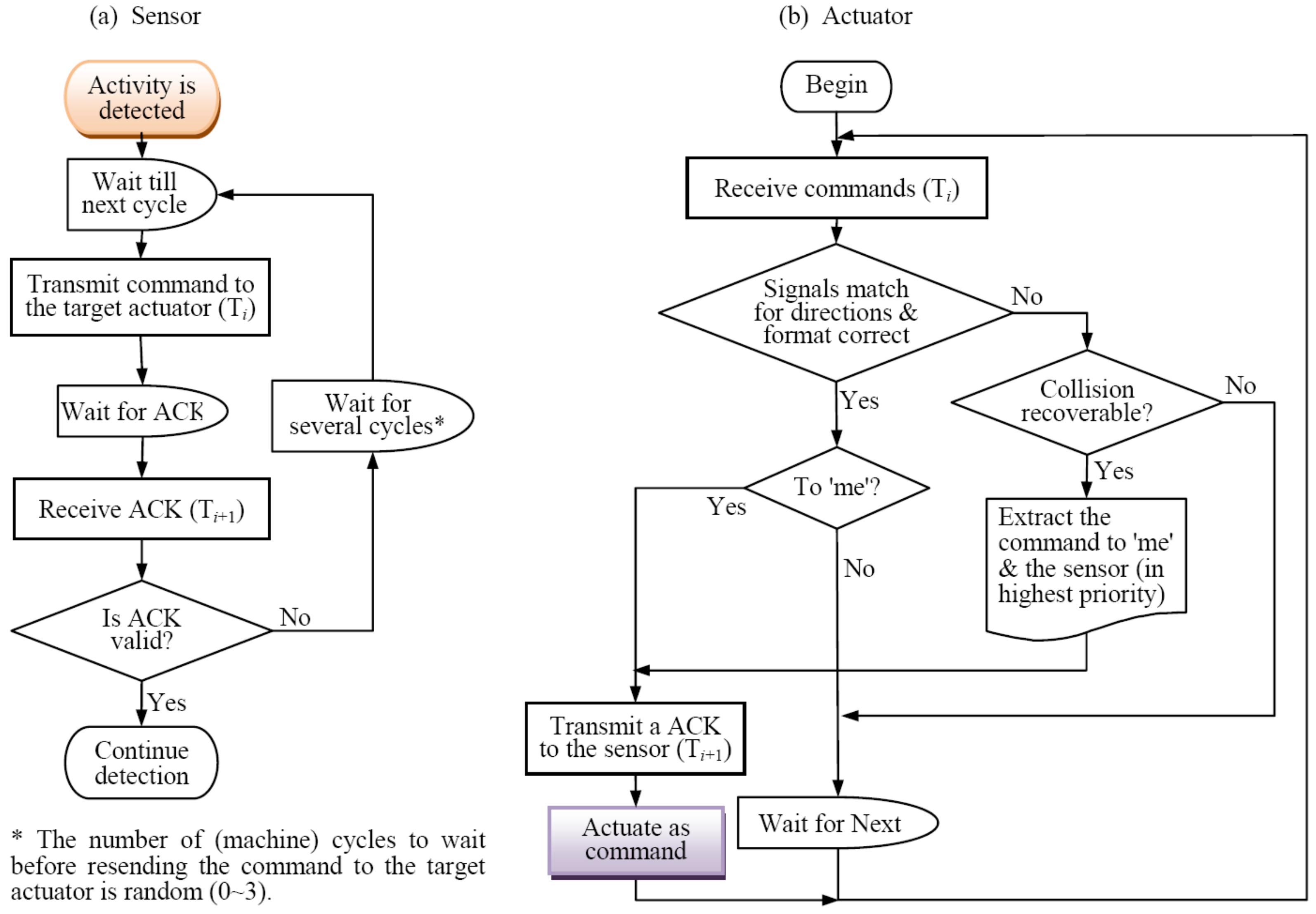}
% where an .eps filename suffix will be assumed under latex,
% and a .pdf suffix will be assumed for pdflatex; or what has been declared
% via \DeclareGraphicsExtensions.
\caption{Flowchart of the synchronous protocol for (a) a sensor (b) a actuator.}
\label{cell_fig11}
\end{figure*}

If an expected activity is probed by a first-layer sensor, in the protocol, this sensor would transmit the command to the target actuator in its recent available transmitting subcycle. The actuator returns an ACK if no signal collision happens during this transmitting subcycle or the command signal can be recovered by the partial space-division technique. The sensor ends this commanding process if the ACK is received; otherwise, the command would be retransmitted after a random delay. The detail of the protocol is shown by the flowchart in Fig. \ref{cell_fig11}. The collision detection is based on the partial space-division and the signal-format verification.

\textbf{\emph{Collision Management Mechanism:}}
In a \emph{in vivo} NIR scenarios, the collision of multiple `0's or multiple `1's would not influence the common recipient, we only need to detect and avoid the collisions of `0' and `1'. To adapt to this scenarios, we proposed a new method of the collision management -- collision detection when muted and exit when colliding (CDWM/EWC). For collision control, when transmitting a bit `0', a first-layer S/A detects the incident power. If the incident power is higher than a given threshold (carrying `1'), the S/A would stop transmit the following bits of this instruction and resend this instruction in the next transmitting subcycle. So, if all other transmitters are visible (i.e., no hidden node exists), this bit collision do not affect the instruction sent by other S/A, and finally, there must be a instruction received correctly. That is, on condition that there is no hidden node, CDWM/EWC is nonblocking. Another feature of CDWM/EWC is its signal priority structure, the instruction with more left `1' has lower probability to collide and exit, so possesses the higher sending priority, which is helpful to introduce new mechanism to promote the system performance.

\begin{figure*}[!t]
\centering
\includegraphics[width=16.0cm]{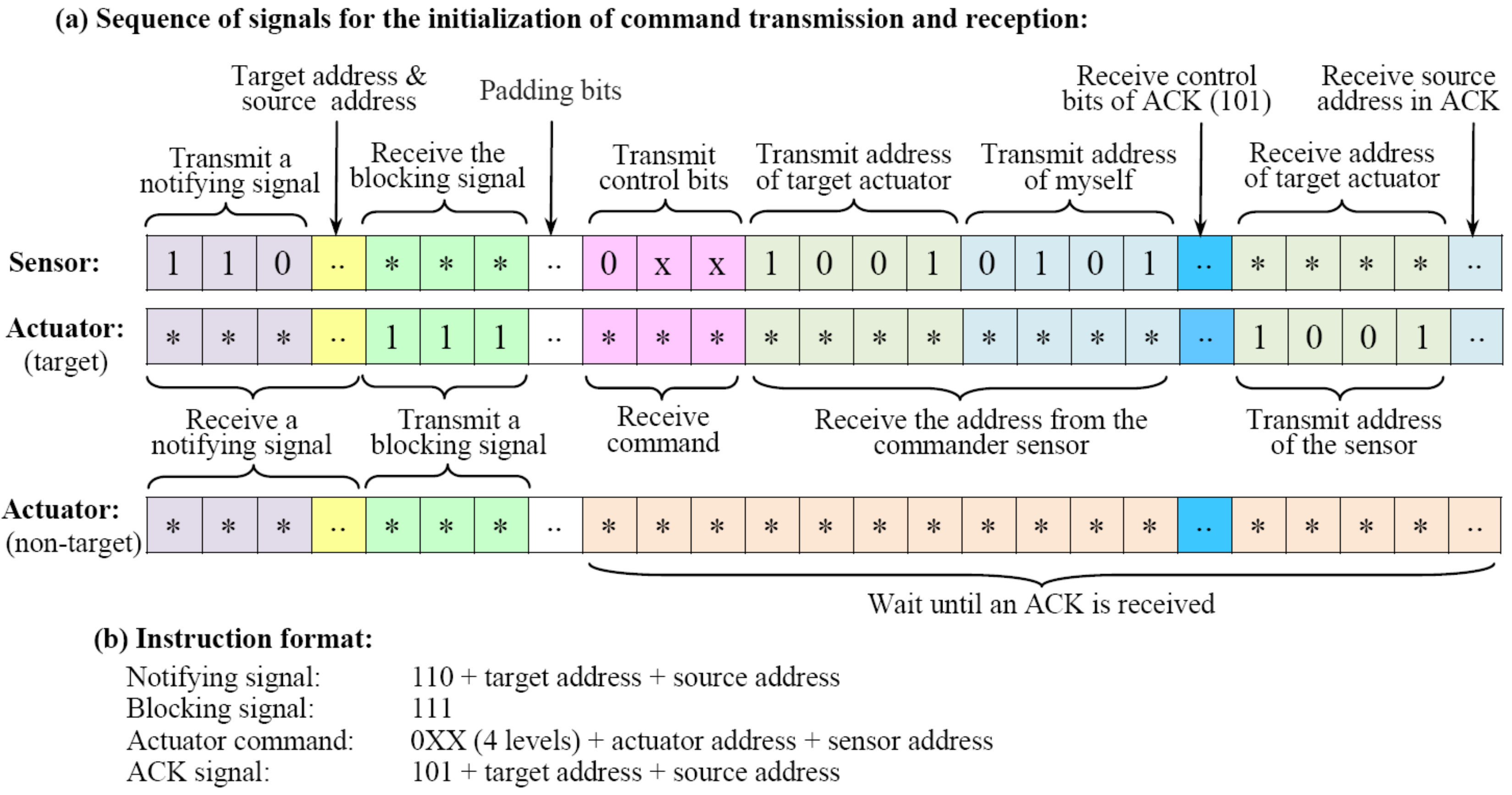}
% where an .eps filename suffix will be assumed under latex,
% and a .pdf suffix will be assumed for pdflatex; or what has been declared
% via \DeclareGraphicsExtensions.
\caption{Format of major instructions in the synchronous protocol.}
\label{cell_fig12}
\end{figure*}

\begin{figure*}[!t]
\centering
\includegraphics[width=16.5cm]{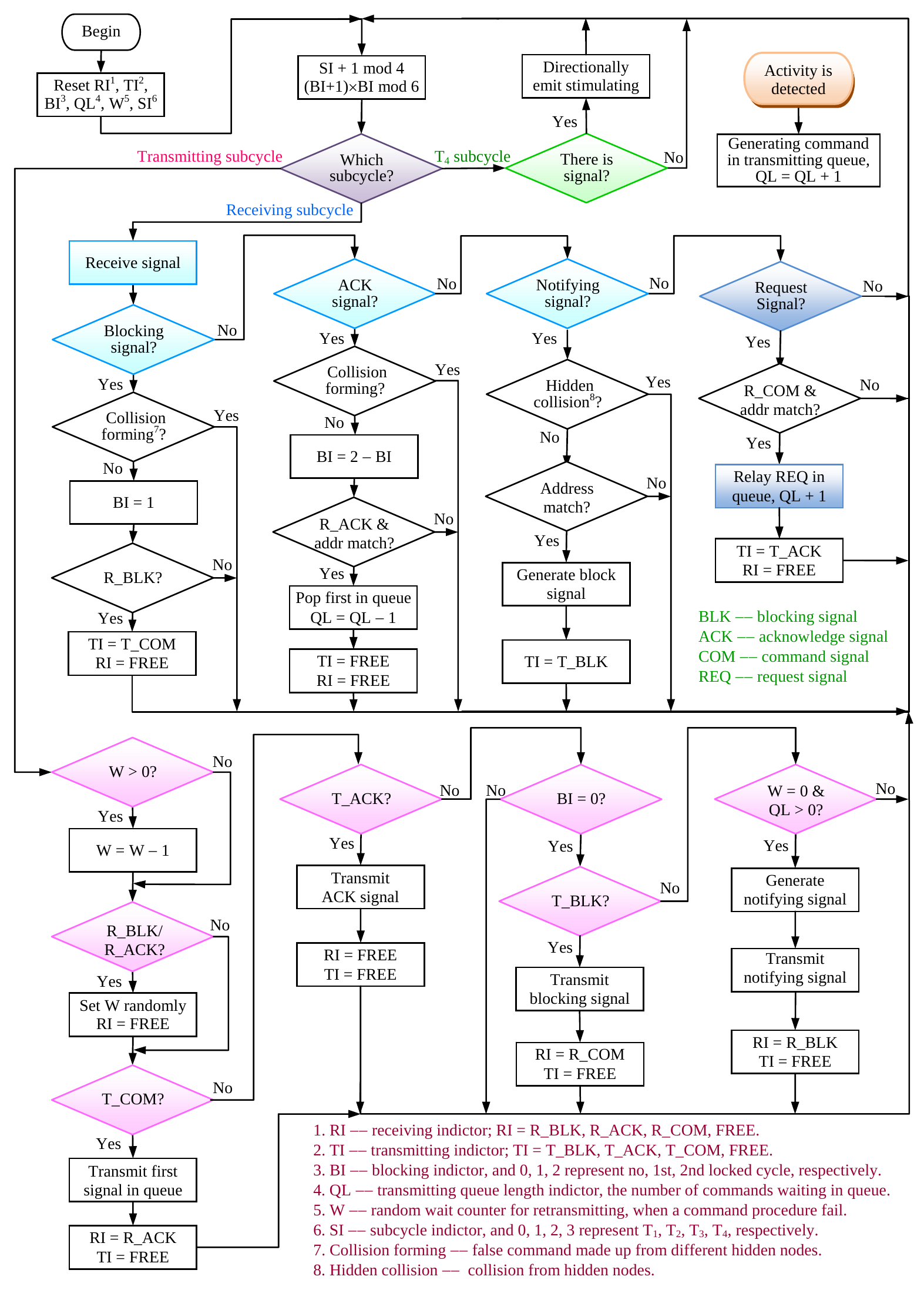}
% where an .eps filename suffix will be assumed under latex,
% and a .pdf suffix will be assumed for pdflatex; or what has been declared
% via \DeclareGraphicsExtensions.
\caption{Working flowchart of the sensors with the collision management mechanism and the stimulating function to the second-layer actuators}
\label{cell_fig13}
\end{figure*}

\begin{figure*}[!t]
\centering
\includegraphics[width=16.5cm]{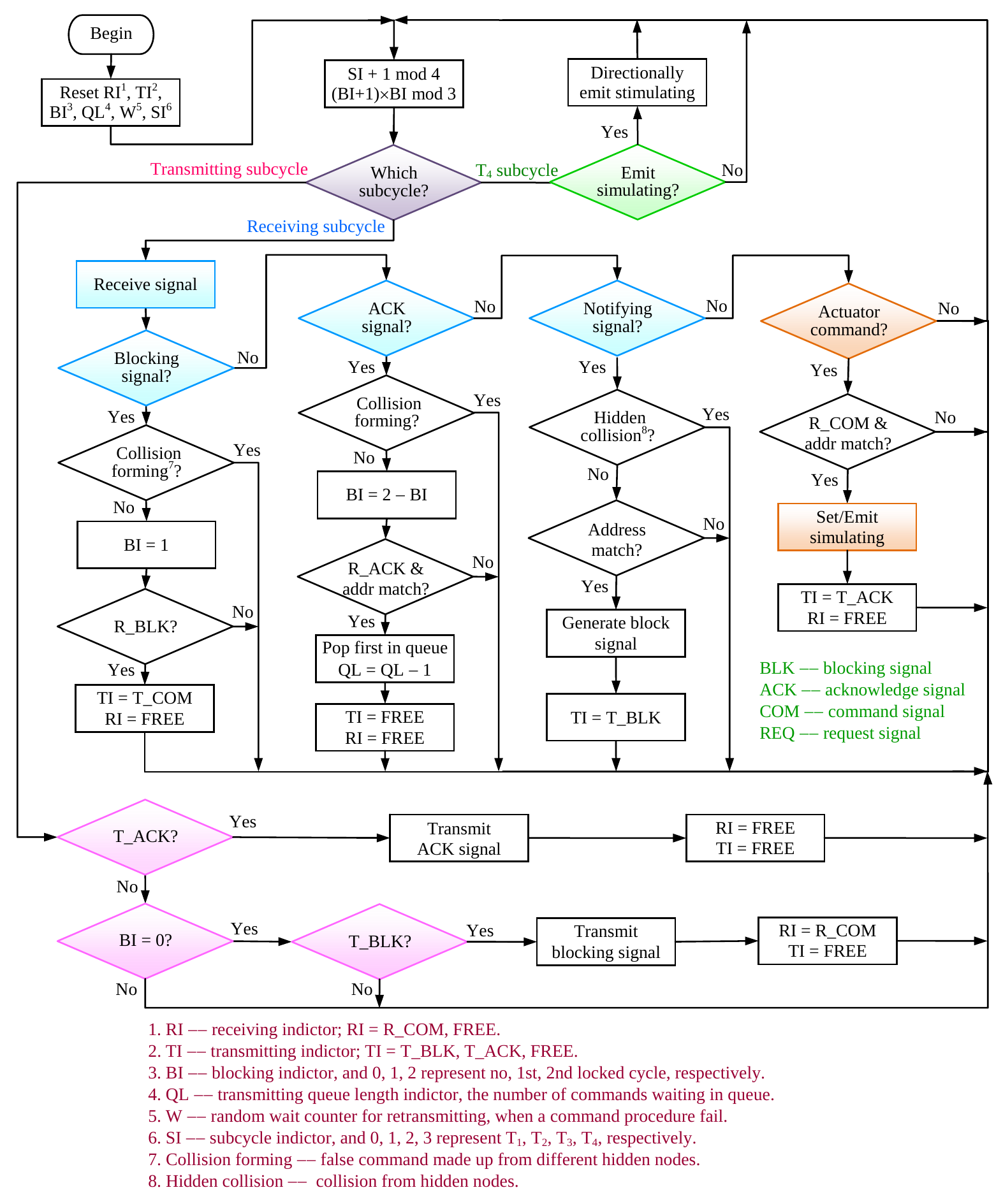}
% where an .eps filename suffix will be assumed under latex,
% and a .pdf suffix will be assumed for pdflatex; or what has been declared
% via \DeclareGraphicsExtensions.
\caption{Working flowchart of the actuators with the collision management mechanism and the stimulating function to the second-layer actuators.}
\label{cell_fig14}
\end{figure*}

But, if there are hidden nodes, the collision from them can not be perceived by a S/A. To solve this problem, we adopt a modified RTS/CTS (Request to Send and Clear to Send) mechanism. When a S/A wants to send a command to its first-layer target node, it first transmits a notifying signal (equivalent RTS signal) to all of its physical recipients. In next transmitting subcycle, the target node, which received the notifying signal without the collision from hidden nodes, broadcasts a blocking signal (equivalent CTS signal) to reserve the share channel. Besides the source node, other recipients of this blocking signal would delay their commands transmitting (if exist) until receiving a reservation relieving signal (ACK signal). This process for command transmission and reception is shown in Fig. \ref{cell_fig12}. In the next transmitting subcycle after received a blocking signal, the source node would transmit the command to the target node. In the following transmitting subcycle, the target node would transmit a ACK signal to acknowledge the command reception. The command process ends normally only when the ACK from the target node is received in the prescribed cycle, otherwise this command process would be repeated after a proper random delay. By directional transmitting and space-division receiving, a receiving node can sense a notifying signal collision from hidden nodes and discard this notifying signal. If no blocking signal received in the next two receiving subcycles, the source node would retransmit the notifying signal after a proper random delay. While the highest sending priority of the blocking signal can ensure it received correctly. The whole protocol is explicitly described by Fig. \ref{cell_fig13} and Fig. \ref{cell_fig14}.

Although some tactics in this protocol is similar to the RTS/CTS used in the WLAN (wireless local-area network) \cite{IEEE11}, but there are mainly two points of contrast: First, in order to deal with the case of dense collision, we consider the collision of the notifying signals (equivalent to RTS), which is however ignored in the WLAN standard 802.11 \cite{IEEE11}. Second, the equivalent signals of RTS and CTS are simplified to several control bits, for example, the blocking signal is `111' (the later address parts is ignored (see Fig. \ref{cell_fig12}). This setting can greatly accelerate the actuation and simplify the signal-processing circuit.

\textbf{\emph{Asynchronous Protocol:}}
An second-layer sensors only has single signal converting function, and has no frame synchronize ability. For example, gold nanorod antennas or SWCNT antennas would generate florescence signals under pulsed laser irradiation, which are asynchronous to frames, despite being bit synchronous. These florescence signals have very less power than the first-layer signal, and last a long time than the instruction cycle once generated. Our asynchronous protocol utilize these features to implement the communications between the asynchronous second-layer sensors and the synchronous first-layer nodes. Since a signals from second-layer sensors has a enough long time, a first-layer S/A only receives the signals from second-layer sensors in T$_4$ subcycle, where no signal from first-layer nodes exists. The second-layer sensors cluster usually on/in tumor cells and generate florescence signals at the same time, thus their signals only need to be identified by space-division, and do not constitute conflicts. So, in T$_4$ subcycle, no signal collision exists. While in T$_1$ $\sim$ T$_3$ subcycles, though there may be the signals from second-layer sensors because of their asynchronism, their power is much less than the signal power from the first-layer nodes and do not influence the correct reception of the signals from the first-layer. Similarly, the second-layer actuators cluster also on/in tumor cells, so that the first-layer node must not identify every second-layer actuator and only need to decide the radiation patterns of the actuator commands. The command to second-layer actuators may be sent in any subcycle, because it is carried on a separate channel and no sensitive to collisions.

\section{Prospective Applications}

This solution has a good application prospect, for example, in cancer treatment. The following we discuss two prospective application examples.

\subsection{Photothermal Cancer Therapy}
In this application example, second-layer nodes adopts SWCNT antennas \cite{Chakravarty2008,Zhou2009} or gold nanorod antennas \cite{Maltzahn2009,Huang2006}, which function as both sensors and actuators, and the first-layer peer-to-peer network consists of only sensors with optical antennas. The energy needed for photothermal therapy is provided by an \emph{ex vivo} near-infrared continuous-wave (CW) laser \cite{Kam2005,Huang2006}.

\begin{figure*}[!t]
\centering
\subfloat[]{\includegraphics[width=8.0cm]{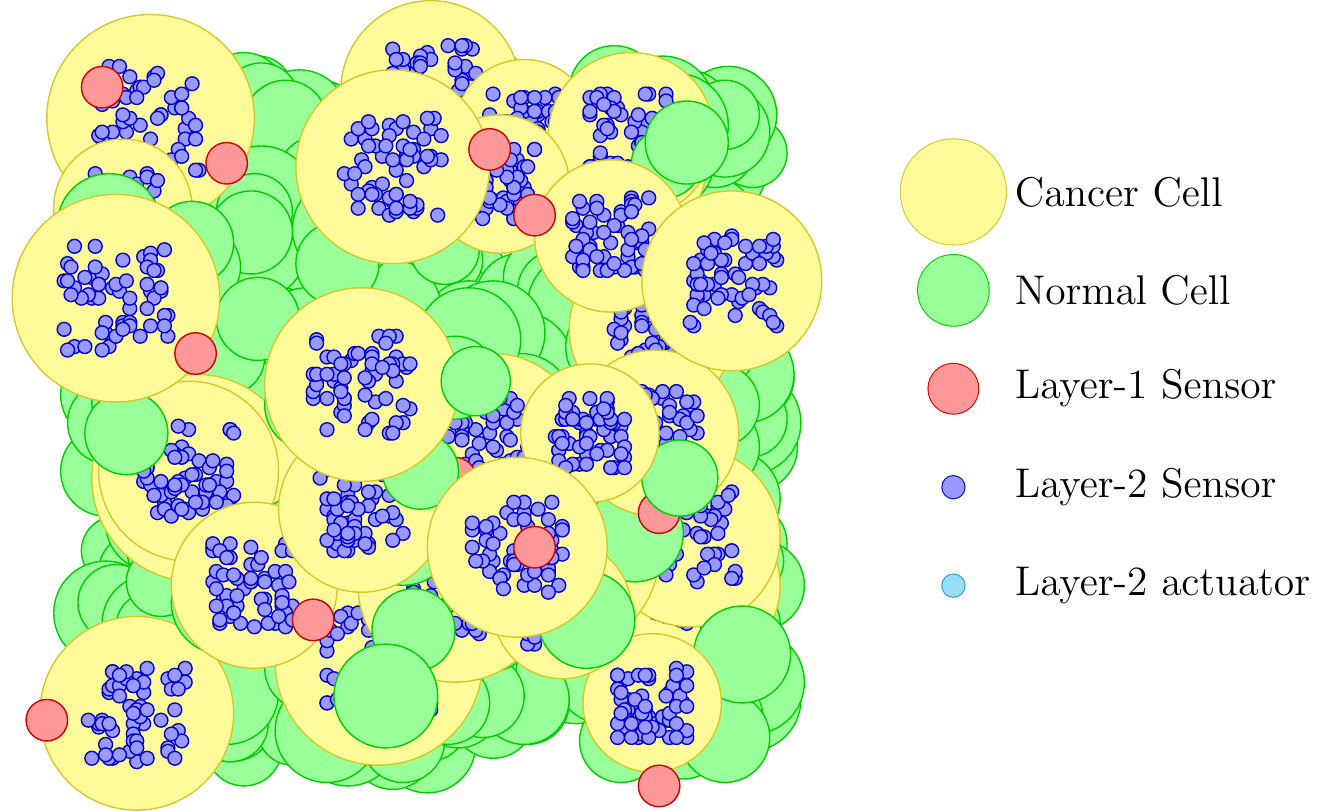}\label{cell_fig15_a}}
\centering
\hspace{1.0cm}
\subfloat[]{\includegraphics[width=5.0cm]{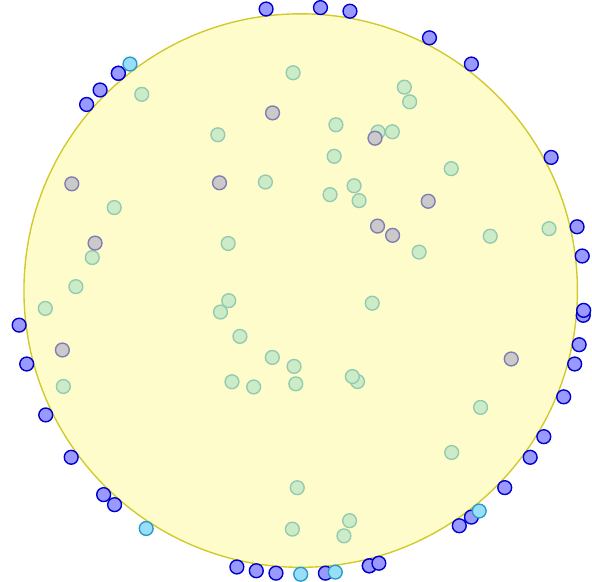}\label{cell_fig15_b}}
\hfil
\caption{Distribution of the second-layer nodes in human body. (a) Clustering distribution of the second-layer sensors/actuators on cancer cells caused by moving to target tumour cells. (b) Clustering distribution of the second-layer sensors/actuators on/in a cancer cell caused by stimulation from first-layer nodes and their own receptor-mediated endocytosis.}
\label{cell_fig15}
\end{figure*}

As second-layer sensors/actuators, the nanoparticles are conjugated to (or coated with) anti-EGFR antibodies \cite{Huang2006} or folate \cite{Kam2005,Zhou2009}, so to automatically locate tumor cells. Going into human body through intravenous injection, these nanoparticles, to target tumor cells, flow along blood vessels and reach, for example, liver, clustering on tumor cells, as shown in Fig. \ref{cell_fig15}(a). Under the irradiation of the carrier pulsed laser, they generate fluorescent signals, informing the nearby first-layer sensor the presence and location of cancer cells. Because very strong clustering effect about cancer cells, nanoparticles straying in the normal cells generate very weaker fluorescent signals, which is much lower than the receiving threshold of the first-layer sensors and are ignored.

On receiving the signal from second-layer sensors, the first-layer sensor will send a request signal to the controller (user interface), notice the area of cancer cells, and require providing or increasing the energy for photothermal therapy. After the request signal of the first-layer sensors is received, the controller will scan and irradiate the designated locations on a certain strategy of increasing the irradiation dose and duration. The second-layer SWCNT antennas or gold nanorod antennas have a very strong absorption to this near-infrared irradiation\cite{Kam2005}, which is converted into heat energy to kill cancer cells. After cancer cells destroyed on the designated locations, their sorption to second-layer SWCNT antennas or gold nanorod antennas will vanish, so that these nanoparticles will dissipate quickly and finally, be eliminated from the body. If a first-layer sensor has not the direct connection with the controller, it can notice the controller through the relay of its adjacent nodes.

\subsection{Anti-Cancer Drug Delivery}
Another example is for anti-cancer drug delivery, which adopts ultra pH-sensitive (UPS) nanoparticles as the second-layer sensors for cancer cell tracking and positioning, and magneto-electric nanoparticles (MENs) as second-layer actuators for anti-cancer drug delivery and controlled release, as shown in Fig. \ref{cell_fig15}(a), \ref{cell_fig15}(b). As in the above example, the biomarker-specific antibodies are attached on the drug-loaded MENs and steer these MENs to the tumor cell membrane \cite{Wang2014}, after they are injected intravenously into human body. Through a similar targeting process, the UPS sensors bind to tumor cell surface receptors and entering inside tumor cells by receptor-mediated endocytosis \cite{Guduru2013}.
The UPS sensors have the ultra pH-sensitive fluorescence emission function with a near-infrared emission range, which can be turned ON/OFF within less than 0.25 pH unit, with a large fluorescence activation ratio (higher than 300-fold). Because the tumour microenvironment has a stronger acidity than normal cells, the UPS sensors stay silent in normal cells, but become activated inside tumors through the sharp pH transition \cite{Zhou2012,Wang2014}. So, after arriving at tumor cells by blood circulation, the UPS sensors will emit fluorescence signal to the nearby first-layer sensor, then the first-layer sensor will send stimulation to MEN actuators. Under the stimulation from the first-layer nodes, the drug-loaded MENs can generate localized electric fields to electroporate the membrane and enter the cancer cells. Next, under a greater stimulation, the MENs will releases anti-cancer drug inside tumor cells \cite{Guduru2013}.

%===========================================================================================================
%                                The Last Section (Conclusion)
%===========================================================================================================
\section{Conclusion}

We have proposed a novel sensing/actuation system featured by cell-level diagnosis and treatment. With the device minimization, especially the development of optical antenna that interacts with the optical field with \cite{Zhao2012,Yu2009}, the design becomes quite meaningful. The system takes the advantage of a common outside synchronous signal realized by laser pulses, as well as a layered Peer-to-Peer network. We have maximally simplified the communication protocol and proposed Collision Detection when Muted and Exit when Colliding (CDWM/EWC), which is adaptive to the high-loss \emph{in-vivo} environment and the characteristics of optical antennas. To specifies topology for each S/A, the learning mode is designed before the working mode of the network. Additionally, we briefly discussed the design of S/A and antenna. The system has promising biomedical applications, such as cancer therapy and anti-cancer drug delivery.

\ifCLASSOPTIONcaptionsoff
  \newpage
\fi

% trigger a \newpage just before the given reference
% number - used to balance the columns on the last page
% adjust value as needed - may need to be readjusted if
% the document is modified later
%\IEEEtriggeratref{8}
% The "triggered" command can be changed if desired:
%\IEEEtriggercmd{\enlargethispage{-5in}}

%===========================================================================================================
%                                                References
%===========================================================================================================
% references section

% can use a bibliography generated by BibTeX as a .bbl file
% BibTeX documentation can be easily obtained at:
% http://www.ctan.org/tex-archive/biblio/bibtex/contrib/doc/
% The IEEEtran BibTeX style support page is at:
% http://www.michaelshell.org/tex/ieeetran/bibtex/
%\bibliographystyle{IEEEtran}
% argument is your BibTeX string definitions and bibliography database(s)
%\bibliography{IEEEabrv,../bib/paper}
%
% <OR> manually copy in the resultant .bbl file
% set second argument of \begin to the number of references
% (used to reserve space for the reference number labels box)

\bibliographystyle{IEEEtran}
\bibliography{IEEEabrv,cell_antenna_nems}

\end{document}